\documentclass{aa}  

\usepackage{newtxtext,newtxmath}

\usepackage{graphicx}	% Including figure files
\usepackage{amsmath}% Advanced maths commands
\usepackage[colorlinks=true,linkcolor=blue,citecolor=blue,urlcolor=blue]{hyperref}
\usepackage{cleveref}  % Optional, enhances cross-referencing
\usepackage{booktabs,tabularx}
\usepackage{threeparttable}
\usepackage{enumitem}
\usepackage{textcomp}
\usepackage{gensymb}
\usepackage{makecell}
\usepackage{bm}
\usepackage{multirow}
\usepackage{verbatim}

\usepackage{pgf}
\usepackage{soul}
\definecolor{lightgreen}{HTML}{B7F774}
\definecolor{lightred}{HTML}{FF6666}
\definecolor{lightorange}{HTML}{FE9A2E}
\sethlcolor{lightgreen}

\defcitealias{Mertens20}{LOFAR20}

\AtBeginDocument{%

}

\setlist[description]{font=\bfseries, leftmargin=0pt}

\begin{document} 

\title{Deeper multi-redshift upper limits on the\\Epoch of Reionization 21-cm signal power spectrum\\from LOFAR between z=8.3 and z=10.1}
\titlerunning{Deeper multi-redshift upper limits on the 21-cm signal power spectrum from LOFAR}
\authorrunning{LOFAR-EoR}

\author{
F. G. Mertens\inst{1,3}\thanks{\email{florent.mertens@obspm.fr}}
\and M. Mevius\inst{2}\thanks{\email{mevius@astron.nl}}
\and L. V. E. Koopmans\inst{3}
\and A. R. Offringa\inst{2,3}
\and S. Zaroubi\inst{4,3}
\and A. Acharya\inst{5}
\and S. A. Brackenhoff\inst{3}
\and E. Ceccotti\inst{3,6}
\and E. Chapman\inst{7}
\and K. Chege\inst{3}
\and B. Ciardi\inst{5}
\and R. Ghara\inst{8}
\and S. Ghosh\inst{3}
\and S. K. Giri\inst{9,10}
\and I. Hothi\inst{1,11}
\and C. Höfer\inst{3}
\and I. T. Iliev\inst{12}
\and V. Jeli\'c\inst{13}
\and Q. Ma\inst{14,15}
\and G. Mellema\inst{16}
\and S. Munshi\inst{3}
\and V. N. Pandey\inst{2}
\and S. Yatawatta\inst{2}
}

\institute{
LUX, Observatoire de Paris, PSL Research University, CNRS, Sorbonne Université, F-75014 Paris, France
\and Astron, PO Box 2, 7990 AA Dwingeloo, The Netherlands
\and Kapteyn Astronomical Institute, University of Groningen, PO Box 800, 9700 AV Groningen, The Netherlands
\and Department of Natural Sciences, The Open University of Israel, 1 University Road, PO Box 808, Ra'anana 4353701, Israel
\and Max-Planck Institute for Astrophysics, Karl-Schwarzschild-Straße 1, 85748 Garching, Germany
\and INAF -- Istituto di Radioastronomia, Via P. Gobetti 101, 40129 Bologna, Italy
\and School of Physics and Astronomy, The University of Nottingham, University Park, Nottingham, NG7 2RD, UK
\and Department of Physical Sciences, Indian Institute of Science Education and Research Kolkata, Mohanpur, WB 741 246, India
\and Van Swinderen Institute for Particle Physics and Gravity, University of Groningen, Nijenborgh 4, 9747 AG Groningen, The Netherlands
\and Nordita, KTH Royal Institute of Technology and Stockholm University, Hannes Alfvéns väg 12, SE-106 91 Stockholm, Sweden
\and Laboratoire de Physique de l'Ecole Normale Supérieure, ENS, Université PSL, CNRS, Sorbonne Université, Université de Paris, F-75005 Paris, France
\and Astronomy Centre, Department of Physics and Astronomy, Pevensey II Building, University of Sussex, Brighton BN1 9QH, U.K.
\and Ru{\dj}er Bo\v{s}kovi\'{c} Institute, Bijeni\v{c}ka cesta 54, 10000 Zagreb, Croatia
\and School of Physics and Electronic Science, Guizhou Normal University, Guiyang 550001, PR China
\and Guizhou Provincial Key Laboratory of Radio Astronomy and Data Processing, Guizhou Normal University, Guiyang 550001, PR China
\and The Oskar Klein Centre, Department of Astronomy, Stockholm University, AlbaNova, SE-10691 Stockholm, Sweden
}

\date{Received XX XX, 20XX}

% \abstract{}{}{}{}{} 
% 5 {} token are mandatory
 
  \abstract
  {We present new upper limits on the 21-cm signal power spectrum from the Epoch of Reionization (EoR), at redshifts $z \approx 10.1, 9.1, \text{ and } 8.3$, based on reprocessed observations from the Low-Frequency Array (LOFAR). The analysis incorporates significant enhancements in calibration methods, sky model subtraction, radio-frequency interference (RFI) mitigation, and an improved signal separation technique using machine learning to develop a physically motivated covariance model for the 21-cm signal. These advancements have markedly reduced previously observed excess power due to residual systematics, bringing the measurements closer to the theoretical thermal noise limit across the entire $k$-space. Using comparable observational data, we achieve a 2 to 4-fold improvement over our previous LOFAR limits, with best upper limits of $\Delta_{21}^2 < (68.7\,\mathrm{mK})^2$ at $k = 0.076\,h\,\mathrm{cMpc}^{-1}$,  $\Delta_{21}^2 < (54.3\,\mathrm{mK})^2$ at $k = 0.076\,h\,\mathrm{cMpc}^{-1}$ and $\Delta_{21}^2 < (65.5\,\mathrm{mK})^2$ at $k = 0.083\,h\,\mathrm{cMpc}^{-1}$ at redshifts $z \approx 10.1, 9.1$, and $8.3$, respectively. These new multi-redshift upper limits provide new constraints that can be used to refine our understanding of the astrophysical processes during the EoR. Comprehensive validation tests, including signal injection, were performed to ensure the robustness of our methods. The remaining excess power is attributed to residual foreground emissions from distant sources, beam model inaccuracies, and low-level RFI. We discuss ongoing and future improvements to the data processing pipeline aimed at further reducing these residuals, thereby enhancing the sensitivity of LOFAR observations in the quest to detect the 21-cm signal from the EoR.}


   \maketitle
%
%-------------------------------------------------------------------
\section{Introduction}

\nolinenumbers

The period following the recombination era ($z \sim 1100$) represents one of the last unexplored frontiers in modern astronomy and cosmology. The epochs starting after the cosmic Dark Ages ($z > 30$), encompass the formation of the first stars and galaxies during the Cosmic Dawn (CD; $15 < z < 30$), and the Epoch of Reionization (EoR; $6 < z < 15$), marking the last major phase transition of the Universe~\citep{Barkana01, Furlanetto06, Loeb13}. Understanding these epochs is crucial for a comprehensive picture of cosmic evolution.

Our knowledge of these early times remains limited, primarily due to the scarcity of detectable luminous sources at such high redshifts. Recent observations, particularly from the James Webb Space Telescope (JWST), have begun to unveil galaxies at redshifts $z > 10$, pushing the boundaries of our observational capabilities~\citep[e.g.,][]{Bouwens23,Harikane23, Atek23, Donnan23, Finkelstein24}. These observations have revealed a surprisingly abundant population of bright galaxies in the early Universe, challenging existing models of galaxy formation and evolution~\citep{BoylanKolchin23,Mason23,ArrabalHaro23}.

Central to exploring these eras is the observation of the redshifted 21-cm line of neutral hydrogen, which promises to chronicle the first billion years of the Universe's evolution. Such observations can yield invaluable insights into cosmic history, the formation of the first luminous structures, and the properties of the sources driving reionization. By tracing the HI gas in the inter-galactic medium (IGM), we can study the cumulative impact of these light sources, not just the brightest ones~\citep[see e.g.,][for reviews]{Ciardi05,Morales10,Pritchard12,Furlanetto16}. These observations may even unveil entirely new physics, such as exotic heating mechanisms or interactions in the dark sector~\citep[e.g.,][]{Barkana18, Fialkov19}.

Current evidence suggests that most of the reionization occurred within $6 \lesssim z \lesssim 10$, as indicated by the Gunn-Peterson trough in high-redshift quasar spectra~\citep[e.g.,][]{Becker01, Fan06,Eilers18} and measurements of the Thomson scattering optical depth of the Cosmic Microwave Background (CMB) radiation~\citep{PlanckCollaboration20}. Recent analyses, particularly those examining IGM damping wing absorption in quasar spectra, imply a rapid evolution of the EoR between $5.5 < z < 7$~\citep[e.g.,][]{Greig17, Davies18,Banados18,Wang20}. Observations of the Ly$\alpha$ forest at $z \sim 6$ suggest that the EoR extended down to $z \sim 5.5$~\citep{Becker15,Bosman18, Keating20,Qin21}. These findings align with CMB constraints, supporting a relatively late EoR with a midpoint around $z \sim 7$~\citep[e.g.,][]{Greig17, Qin20}. Nonetheless, much remains to be understood about the EoR, and the observation of the 21-cm signal from this epoch could provide critical insights into these formative periods of the Universe.

Large low-frequency radio telescopes such as LOFAR\footnote{LOw-Frequency ARray, \url{http://www.lofar.org}}~\citep{Haarlem13}, MWA\footnote{Murchison Widefield Array, \url{http://www.mwatelescope.org}}~\citep{Tingay13}, NenuFAR\footnote{New Extension in Nançay Upgrading LOFAR, \url{https://nenufar.obs-nancay.fr/en/homepage-en/}}~\citep{zarka:hal-04056720}, HERA\footnote{Hydrogen Epoch of Reionization Array, \url{https://reionization.org/}}~\citep{Deboer17}, and GMRT\footnote{Giant Metrewave Radio Telescope, \url{http://www.gmrt.ncra.tifr.res.in}}~\citep{Gupta17} are at the forefront of the search for the 21-cm signal across a broad range of redshifts. Initially, these experiments aim for a statistical detection of the 21-cm fluctuations. These efforts are also crucial for the success of the forthcoming SKAO\footnote{Square Kilometre Array Observatory, \url{https://www.skao.int}}, a next-generation radio telescope with unmatched sensitivity and with the potential to obtain images of these epochs~\citep{Koopmans15a,Mellema15}. Despite significant challenges, these instruments have set impressive upper limits on the 21-cm signal power spectra, but they have yet to achieve a detection. Recent efforts include those by~\cite{Trott20}, who reported a 2-$\sigma$ upper limit of $\Delta_{21}^2 < (43.9~\mathrm{mK})^2$ at $z \approx 6.5$ and $k = 0.15~h\,\mathrm{cMpc^{-1}}$ with the MWA, using 298 h of carefully selected data, and the HERA team, who recently reported a 2-$\sigma$ upper limit of $\Delta_{21}^2 < (21.4~\mathrm{mK})^2$ at $z \approx 7.9$ and $k = 0.34~h\,\mathrm{cMpc^{-1}}$, and $\Delta_{21}^2 < (59.1~\mathrm{mK})^2$ at $z \approx 10.4$ and $k = 0.36~h\,\mathrm{cMpc^{-1}}$using 94 nights of observations~\citep{HERACollaboration23}. The LOFAR-EoR Key Science Project has also made significant progress. We published a first upper limit based on 13 hours of data~\citep{Patil17}. This was later improved to a 2-$\sigma$ upper limit at $z \approx 9.1$ of $\Delta^2_{21} < (72.86\, \mathrm{mK})^2$ at $k = 0.075~h\,\mathrm{cMpc^{-1}}$, using 141 hours of data~\citep[][hereafter \citetalias{Mertens20}]{Mertens20}.

Detecting the 21-cm signal is extremely challenging due to the difficulty of extracting this faint signal buried beneath astrophysical foregrounds that are many orders of magnitude brighter and contaminated by numerous systematics. At the low radio-frequencies targeted by 21-cm signal observations, the emission from the Milky Way and other extra-galactic sources dominates the sky. The emission of these foregrounds varies smoothly with frequency, which can be used to differentiate it from the rapidly fluctuating 21-cm signal~\citep{Jelic08}. However, the frequency-dependent response of the radio telescopes introduces structure to the otherwise spectrally-smooth foregrounds, causing so-called `mode-mixing'~\citep{Morales12}. Most chromatic effects are confined inside a wedge-like shape in $k$-space~\citep{Datta10, Trott12, Vedantham12, Liu14a}. High-precision calibration is essential, as errors can be introduced by calibration with an incomplete or incorrect sky model~\citep{Patil16,Ewall-Wice17,Barry16}, incorrect bandpass calibration, cable reflections~\citep{Beardsley16}, as well as chromatic errors due to leakage from the polarized sky into Stokes-I~\citep{Jelic10, Spinelli18}, ionospheric disturbances~\citep{Koopmans10,Vedantham16,Jordan17,Mevius16,Brackenhoff24}, incorrect primary-beam models~\citep{Gehlot21,Chokshi24} or gridding errors~\citep{Offringa19}. Multi-path propagation, mutual coupling~\citep{Kern20,Kolopanis23} and residual radio-frequency interference (RFI)~\citep{Offringa19a,Wilensky19} must also be corrected or mitigated with great precision to detect the redshifted 21-cm signal.

Various observing and analysis strategies have been implemented by different teams, to optimise the telescope's capabilities and ensure the successful detection of the 21-cm line. The strategy of the LOFAR-EoR Key Science Project involves combining thousands of observing hours on a deep field, calibrated to high precision with a deep and extended sky model of the field using a sky-based calibration scheme. We also pursue a foreground removal strategy to model and remove foreground contaminants, aiming to probe the 21-cm signal both outside and inside the foreground wedge, enhancing sensitivity and accessing larger scales. This effort requires a comprehensive sky model. In this work we focus on one of the deep fields studied by the LOFAR-EoR Key Science Project, namely the North Celestial Pole (NCP). The model~\citep{Yatawatta13, Patil17} of this field currently consists of almost thirty thousand components. This model is used for solving station gains in multiple directions using the \textsc{Sagecal-CO} code~\citep{Yatawatta16} and subsequently removing these components with their direction-dependent instrumental response functions. Residual foregrounds are then statistically separated from the 21-cm signal using Gaussian Process Regression~\citep[GPR,][]{Mertens18a,Mertens24}.

This strategy was implemented by~\citetalias{Mertens20}, where, despite setting scientifically interesting upper limits and discarding some extreme models~\citep{Ghara20, Greig21, Mondal20}, the results were still above the theoretical limits expected if thermal noise dominated. Over the past four years, many potential sources of this excess have been investigated, including residual foreground emissions from off-centre sources, chromatic direction-independent (DI) and direction-dependent (DD) calibration errors, low-level RFI, and ionospheric disturbances. Detailed analyses by~\cite{Gan22} showed that the excess variance was not strongly correlated with gain variance or ionospheric conditions, suggesting that neither calibration errors nor ionospheric effects are the primary contributors. This finding was corroborated by~\cite{Gan23}, who found no significant difference in foreground removal between two calibration algorithms, and by~\cite{Brackenhoff24}, who demonstrated that ionospheric impact on cylindrically averaged power spectra is confined to the wedge and can be effectively modelled and removed using current techniques. However,~\cite{Gan22} indicated that the excess variance might be related to bright, distant sources such as Cassiopeia\,A (Cas\,A) and Cygnus\,A. Additionally,~\cite{Hothi21} found that the GPR method was optimal compared to alternatives, although~\cite{Kern21} highlighted the need for a more physically motivated approach to GPR to reduce the risk of bias and signal loss.

Although we have not yet definitively identified the cause of the observed excess power in our data, these recent analyses have provided valuable insights into potential contributing factors, leading to significant enhancements in our processing pipeline. These include improvements in both DI and DD-calibration, more effective RFI flagging, and enhanced residual foreground removal methods. In particular, the later benefited from several recent works: \cite{Mertens24} introduced the concept of learned kernels, which allows for a parametrized covariance function to be derived from simulated datasets, ensuring a more physically motivated model and significantly improving component separation. This new ML-GPR method was first tested on LOFAR simulations~\citep{Acharya24} and then applied to LOFAR data in~\cite{Acharya24a}. 
% The updated sky model also includes a new model for 3C\,61.1, a relatively bright source in the NCP field.

In this publication, we present improved multi-redshift 21-cm power spectrum upper limits from the LOFAR-EoR Key Science Project. These new upper limits are based on a similar set of observations used in our~\citetalias{Mertens20} publication (about 140h of data) but include a broader frequency range (122--159 MHz). This broader range allows us to set upper limits at redshifts centred on $z \approx 10.1, 9.1, \text{ and } 8.3$. 

Our observational strategy is detailed in~\autoref{sec:observation}, with the processing and analysis methods described in~\autoref{sec:data_reduction}. A new upper limit on the 21-cm signal power spectra is presented in~\autoref{sec:results}, and the validation procedure is explained in~\autoref{sec:validations}. Finally, we discuss our results and their implications in~\autoref{sec:discussion}. Throughout this paper we use a $\Lambda$CDM cosmology consistent with the Planck 2015 results~\citep{PlanckCollaboration16}.

\section{Observations}
\label{sec:observation}

\begin{table*}
\centering
\caption{List of all observation nights analysed in this work, detailing date, time, duration, and noise statistics for each night. Note that the SEFD measurements have been updated compared to the one published in \citetalias{Mertens20}, following an improved methodology. The corresponding LOFAR cycles have been corrected with respect to \citetalias{Mertens20} as well.}
\begin{threeparttable}
\label{table:observations}
\label{tab:all_nights}
\begin{tabular}{lrrrrrrccc}
\toprule
\multirow{2}{*}{\makecell{Night ID\\\,}} & \multirow{2}{*}{\makecell{LOFAR\\Cycle}} & \multirow{2}{*}{\makecell{UTC observing start\\date and time}} & \multirow{2}{*}{\makecell{LST$^a$ starting\\time [hour]}} & \multirow{2}{*}{\makecell{Duration [hour]\\\,}} & \multirow{2}{*}{\makecell{SEFD$^b$ estimate \\\,[Jy]}} & \multicolumn{3}{c}{Redshift selection$^c$}\\
\cmidrule(lr){7-9}
& & & & & & \makecell{8.3} & \makecell{9.1} & \makecell{10.1} \\
\midrule
L80847 & 0 & 2012-12-31 15:33:06 & 22.7 & 16.0 & 4409 & $\checkmark$ & & $\checkmark$ \\
L80850 & 0 & 2012-12-24 15:30:06 & 22.2 & 16.0 & 4448 & & $\checkmark$ & $\checkmark$ \\
L86762 & 0 & 2013-02-06 17:20:06 & 2.9 & 13.0 & 4349 & $\checkmark$ & $\checkmark$ & \\
L90490 & 0 & 2013-02-11 17:20:06 & 3.2 & 13.0 & 4642 & $\checkmark$ & $\checkmark$ & $\checkmark$ \\
L196421 & 1 & 2013-12-27 15:48:38 & 22.7 & 15.5 & 4292 & $\checkmark$ & & $\checkmark$ \\
L203277 & 1 & 2014-02-17 17:14:20 & 3.5 & 13.1 & 3935 & & $\checkmark$ & \\
L205861 & 1 & 2014-03-06 17:46:30 & 5.2 & 11.9 & 3917 & & $\checkmark$ & $\checkmark$ \\
L246291 & 2 & 2014-10-25 16:42:14 & 19.4 & 13.2 & 4261 & $\checkmark$ & $\checkmark$ & $\checkmark$ \\
L246297 & 2 & 2014-10-23 16:46:30 & 19.3 & 13.0 & 4309 & $\checkmark$ & $\checkmark$ & $\checkmark$ \\
L246309 & 2 & 2014-10-16 17:01:41 & 19.1 & 12.6 & 4402 & $\checkmark$ & $\checkmark$ & $\checkmark$ \\
L253987 & 3 & 2014-12-05 15:44:35 & 21.1 & 15.3 & 4105 & $\checkmark$ & $\checkmark$ & $\checkmark$ \\
L254116 & 3 & 2014-12-10 15:42:54 & 21.4 & 15.4 & 4515 & $\checkmark$ & $\checkmark$ & \\
L254865 & 3 & 2014-12-23 15:45:36 & 22.3 & 15.5 & 4156 & $\checkmark$ & & \\
L254871 & 3 & 2014-12-20 15:44:04 & 22.1 & 15.5 & 4170 & & & $\checkmark$ \\
\bottomrule
\end{tabular}
\begin{tablenotes}
  \small
  \item[a] Local Sidereal Time.
  \item[b] System Equivalent Flux Density, estimated from time-differenced visibility in the frequency range 134-147 MHz.
  \item[c] Observation selected for analysis in redshift bin 8.3, 9.1, and/or 10.1.
\end{tablenotes}
\end{threeparttable}
\end{table*}

The LOFAR radio telescope comprises 24 core stations distributed within a 2 km diameter, 14 remote stations across the Netherlands, providing a maximum baseline length of approximately 100 km, and an increasing number of international stations across Europe~\citep{Haarlem13}. The LOFAR-EoR observations utilise the High Band Antennas (HBA), operating at frequencies between 110 and 189 MHz, and target two primary fields: the North Celestial Pole (NCP) and the bright compact radio source 3C\,196~\citep{Bruyn12}.

We reanalysed 12 NCP observations from LOFAR Cycles 0 to 3, which were previously used in~\citetalias{Mertens20}, along with two additional nights from Cycles 1 and 2. This amounts to about 200 hours of observation. The NCP is particularly advantageous for EoR studies due to its year-round continuous visibility, making it a key focus for deep field observations. These observations employed all core stations (in split mode, effectively providing 48 stations) along with remote stations. The observational setup remained nearly identical to that described in~\citetalias{Mertens20}. Data were recorded with a frequency range of 115 to 189 MHz, a spectral resolution of 3.05 kHz (resulting in 64 channels per sub-band of 195.3 kHz), and a temporal resolution of 2 seconds. Key differences from the~\citetalias{Mertens20} analysis include the processing of two additional frequency bands -- 122--134 MHz and 147--159 MHz -- alongside the previously used 134--147 MHz band, as well as the inclusion of two additional nights of observations. All frequency bands were reprocessed using an updated processing pipeline, which is detailed in the subsequent sections. A summary of all observations is provided in Table~\ref{table:observations}.

\section{Data Reduction}
\label{sec:data_reduction}

\begin{figure*}
\includegraphics{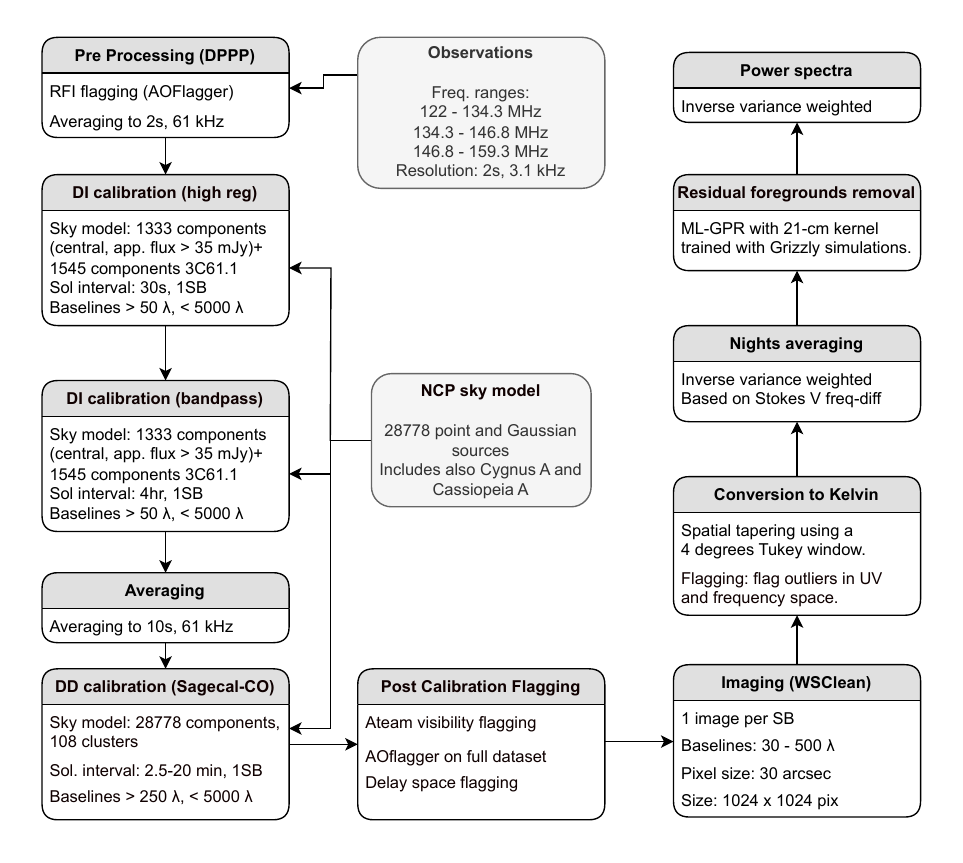}
\caption{\label{fig:lofar_hba_ncp_processing} The LOFAR-EoR HBA processing pipeline, describing the steps required to reduce the raw observed visibilities to the 21-cm signal power spectra.}
\end{figure*}

The LOFAR-EoR data processing pipeline has been iteratively developed and generally follows the strategy described by~\cite{Patil17} and~\citetalias{Mertens20}. It comprises the following key steps: (i) Pre-processing, where visibility averaging and RFI excision are performed; (ii) Calibration, involving a direction-independent (DI) calibration scheme using a sky-based approach; (iii) Sky-model source subtraction, employing direction-dependent (DD) calibration for accurate source removal; (iv) Post-calibration flagging, which includes further RFI excision; (v) Imaging, converting visibilities into image cubes in units of Kelvin; (vi) Combination of nights, where data from multiple nights are combined using an inverse variance-weighted method; and finally, (vii) Residual foreground removal, which models the gridded observed data as a sum of multiple components: foregrounds, excess emissions, and the 21-cm signal. The foreground component is removed, and the residual forms the basis for our upper limit on the 21-cm power spectrum.

This study introduces several improvements, particularly in the calibration steps, sky model subtraction, and post-calibration flagging strategy. A new method for residual foreground subtraction was also implemented. Figure~\ref{fig:lofar_hba_ncp_processing} shows an overview of the LOFAR-EoR data processing pipeline. All data processing was conducted on the `Dawn' compute cluster, equipped with 48 × 32 hyperthreaded compute cores and 124 Nvidia K40 GPUs, located at the Centre for Information Technology of the University of Groningen.

% RFI flagging and averaging
\subsection{Pre-processing}

The raw LOFAR NCP data are initially integrated with a time resolution of 2 seconds and 64 channels per sub-band. After applying RFI flagging using AOFlagger~\citep{Offringa12} and excluding the outer 2 channels on both sides of the band, the data are averaged to 15 channels of 12.2 kHz per sub-band before archiving in the LOFAR Long Term Archive (LTA). As described in \citetalias{Mertens20}, a second AOFlagger step is performed before further averaging to three channels of 61 kHz width. Intra-station baselines, affected by crosstalk due to shared electronics cabinets, are also fully flagged. To stabilize the initial calibration, the raw data are pre-scaled such that the visibility amplitudes are between 1 and 10. This step makes sure that the initial gains, initialized with an amplitude of 1 and phase of zero, are within a factor of 10 of the final optimised gains. The resulting data product, averaged over three channels with a resolution of 2 seconds, is then used in the initial calibration step.

\subsection{Calibration}

\begin{figure}
    \includegraphics{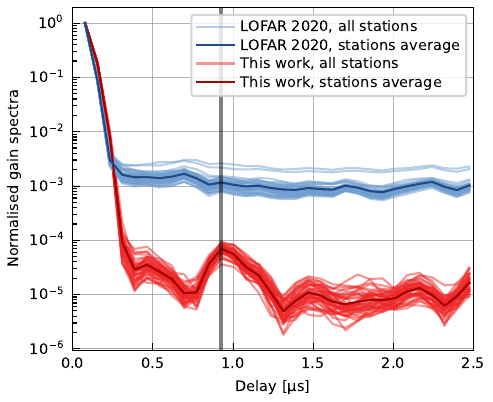}
    \caption{Normalized gain-spectra of the DI-calibration for frequency range 134-147 MHz, observation L254871. The one-step unregularized calibration (`LOFAR 2020', blue lines) is compared with the two-steps calibration (`This work', red lines). The delay corresponding to reflections in the 115-meter cables of LOFAR is denoted by a grey line, where a spectral feature is also observed, as expected.}
    \label{fig:di_gain_spectra}
\end{figure}
For calibration, we used a sky model similar to that described in~\citetalias{Mertens20}, with the exception of modified models for the three brightest sources (3C\,61.1, Cas\,A, and Cygnus\,A) and a reduced number of components and clusters. Details of these modifications are provided below. The intensity scale of the the sky-model is set by NVSS~J011732+892848 (RA~01h\,17m\,33s, Dec~89$\degree$\,28'\,49'' in J2000), as was the case in \citetalias{Mertens20}.

The spectral variability of the bright 3C\,61.1 source near the first null of the beam, would dominate the calibration solutions in the rest of the field. Therefore, similar to \citetalias{Mertens20}, the `direction-independent' calibration step treats the  3C\,61.1 direction separately. It solves simultaneously for the station gains in two directions and then corrects the visibilities with the gain solutions of the central field.  We solve for a 2$\times$2 complex gain (i.e. Jones) matrix per station, using a $\sim 1300$ component model of the central field as well as an updated model consisting of $1545$ clean components for 3C\,61.1 \citep{Ceccotti23}. The model consists of intrinsic flux density values, and the station beam response is modelled from the geometrical delays using the positions of the individual dipoles (i.e. the array factor). Tiles that were flagged during the observations are taken into account in the beam response model, but individual malfunctioning dipoles are not. In this way the dipole response including any cross coupling effects are absorbed in the gain solutions and thus applied when using the gains to correct the visibilities. 

\textsc{Sagecal-CO}~\citep{Yatawatta16} is used to calibrate our observations, which allows us to regularize the spectral behaviour of the gain solutions to approach a smooth curve by a regularisation prior. We use a third-order Bernstein polynomial over the 13-MHz bandwidth, treating each redshift bin separately. Variation of the \textsc{Sagecal-CO} regularisation parameter determines how fast the solutions converge to this smooth curve. In \citetalias{Mertens20} almost full spectral freedom was allowed during DI-calibration by setting a low regularisation parameter and only relatively few iterations. It is known that spectral errors in the gains that are applied when correcting the visibilities can introduce unwanted features in the final power spectrum that cannot be removed in later steps in the processing. Unmodelled sky components can for example be the cause of such errors in the gains~\citep[e.g.][]{Barry16}. Following \citet{Mevius22}, a high spectral regularisation of the gains is desired. However, small spectral features, such as cable reflections, with a typical frequency scale of $\sim 1$ MHz for LOFAR HBA data, require solutions with high spectral resolution. Those features, however, are expected to be fairly constant in time and direction. We therefore adopt a two-step approach:
% \begin{enumerate}
\begin{enumerate}[leftmargin=*,label=(\roman*)]
    \setlength\itemsep{0.5em}
    \setlength\labelwidth{0pt}
    % \setlength\itemsep{0.25em}
    % \item[(1) The foregrounds ---]
    \item \textbf{DI-calibration with high spectral regularisation ---} First, DI spectrally smooth solutions are fitted using \textsc{Sagecal-CO} with a high-regularisation parameter and a solution time interval of 30s.

    \item \textbf{Bandpass Calibration ---} Subsequently, a full bandpass calibration is performed with a single solution per sub-band and a time interval of 4 hours. This interval is chosen due to computational limitations, although ideally, the entire observation duration would be used.
\end{enumerate}
Note that a simultaneous solve for long and short term solution intervals would also be computationally impossible. Therefore, this iterative approach was chosen. The preferred order of the two-step approach, first high-regularisation then bandpass calibration, is based on experiments on actual data. Convergence of the solutions was most stable this way and experiments showed no further improvement after a second iteration of high-regularisation calibration.

% Those features, however, are expected to be fairly constant in time and direction. We therefore have chosen a two-step approach. First, DI spectrally smooth solutions are fitted using \textsc{Sagecal-CO} with a high-regularisation parameter and a solution time interval of 30s. After that, a full bandpass calibration is done with a single solution per sub-band and a time interval of 4 hours. This 4-hour interval is chosen due to computational limitations, although ideally, the entire observation duration would be used. Note that a simultaneous solve for long and short term solution intervals would also be computationally impossible. Therefore, this iterative approach was chosen. The preferred order of the two-step approach, first high-regularisation then bandpass calibration, is based on experiments on actual data. Convergence of the solutions was most stable this way and experiments showed no further improvement after a second iteration of high-regularisation calibration. 

The  sky model is more accurate on the shorter baselines, where it is less affected by ionospheric errors~\citep[e.g.][]{Vedantham16,Mevius16,Ewall-Wice17,Brackenhoff24}. Therefore, we decided to limit the baseline range of the visibilities  used in the DI-calibration to 50 -- 5000 $\lambda$. This choice includes all LOFAR core baselines and is based on experiments testing the final power spectrum of a single night of observation with various baseline cuts and solution time intervals. Importantly, this baseline length is comparable to or smaller than the typical ionospheric diffractive scale at our observing frequency~\citep{Mevius16}, minimising the impact of ionospheric phase fluctuations. A more rigorous analysis of the optimal baseline selection using a full simulation of the processing pipeline will be the topic of a follow up paper. 

\autoref{fig:di_gain_spectra} shows the gain spectra of the DI-calibration solution for a typical observation night. It demonstrates a significant reduction in gain errors by 2 to 3 orders of magnitude compared to~\citetalias{Mertens20}, while preserving the correction of spectrally non-smooth but time-stable features in the signal processing chain (e.g., cable reflections), which are properly accounted for.

\subsection{Sky-model sources subtraction}

In the second calibration step no corrections are applied to the data. Instead, an extensive model, consisting of 28,778 components, is subtracted from the visibilities after multiplication with the fitted DD-gains. As was the case in \citetalias{Mertens20}, we fit the DD-gains on a different set of baselines than those used in the final power spectrum. We calibrate our data on the baselines between $250$ and $5000~\lambda$, while the final power spectrum uses the baselines below $250~\lambda$. The reason for this is two-fold: first, this method ensures that the 21-cm signal of interest is not suppressed whilst subtracting the gain-corrected model \citep{Sardarabadi19, Mevius22} and secondly, our sky model does not include the large diffuse structures which mainly affect the shorter baselines that are excluded from the gain calibration. The DD-calibration closely follows that of \citetalias{Mertens20} with a few alterations highlighted here. 

\begin{figure}
    \includegraphics{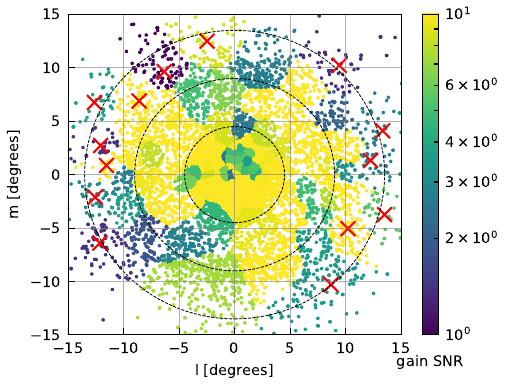}
    \caption{Signal to noise ratio (S/N) of the time-differenced solutions for all $\sim 28,000$ sky components of \protect\citetalias{Mertens20} in real data. Clusters of components can be recognised since they share the same gain solutions. In total, there are 120 clusters in the image. The S/N is similar for most clusters, but showing a slight gradient to lower S/N values away from the phase centre. The components in the outer 14 clusters (with a flux weighted mean position more than 10 degrees away from the phase centre), indicated with red crosses, have been removed from the model used in the current analysis. }
    \label{fig:gain_cluster_variance}
\end{figure}

\citetalias{Mertens20} used a 28,773 component model divided into 122 separate clusters, where the $2\times2$ complex station gains in the direction of each individual cluster were solved for simultaneously using \textsc{Sagecal-CO}. In this work, we modify that model in several ways. Firstly, we found that residuals from the bright A-team sources (Cas\,A and Cygnus\,A) significantly contributed to excess variance in the power spectrum~\citep{Gan22}. We therefore updated their model, replacing shapelet components with a simpler mix of point sources and Gaussians, which improved results, although the exact reason for this improvement is still under investigation \citep{Ceccotti25}. Secondly, clusters outside the first station beam null showed higher variance in the solutions, likely due to rapid beam-related gain fluctuations that cannot be accurately captured by the spectrally-smooth gains used in \textsc{Sagecal-CO}. In such cases, including these clusters can increase, rather than reduce, the overall variance. By excluding 14 outer clusters from the DD-calibration (see \autoref{fig:gain_cluster_variance}), we achieved a marked improvement in the final power spectrum. The updated model now contains 28,778 components in 108 clusters.

\begin{figure}
    \includegraphics{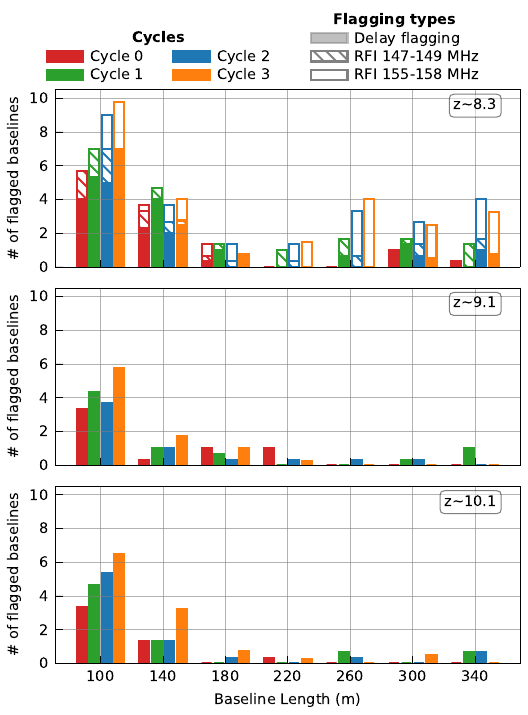}
    \caption{Number of flagged baselines by some of the post-calibration visibility flagger, grouped by baseline length bins and categorised by different sources of contamination. Short baselines are predominantly affected, although certain frequency bands (147--149 MHz and 155--159 MHz) also show contamination on longer baselines}
    \label{fig:vis_flagger}
\end{figure}

\begin{figure*}
    \includegraphics{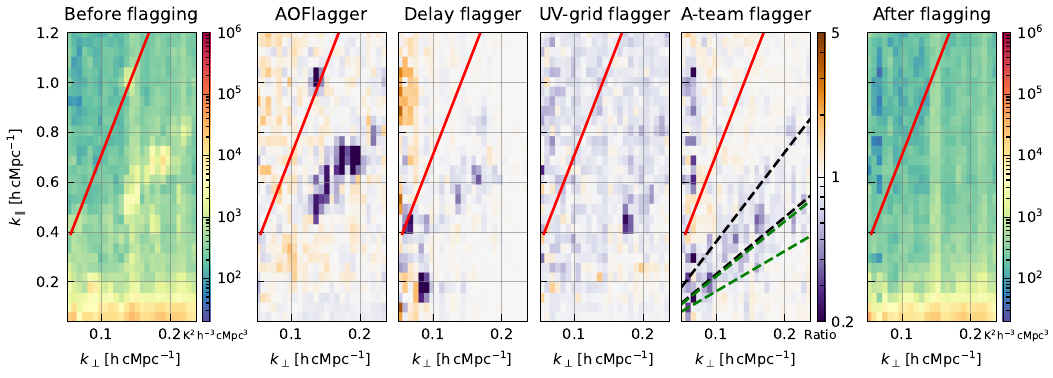}
    \caption{Effect of post-calibration flagging steps on the 2D power spectra for observation L246309 in the frequency range 134–147 MHz. The left panel shows the power spectra before any calibration. The four middle panels display the ratio of power spectra before and after each successive flagging step, illustrating the progressive reduction of contamination. The four steps, shown in order, are: (1) post-calibration wide-bandwidth AOFlagger; (2) baseline flagging of delay-spectrum outliers above the horizon limit; (3) post-gridding flagging based on outlier detection in channels and uv-cells; and (4) post-gridding flagging of uv-cells contaminated by A-team sidelobes. The right panel shows the power spectra after all steps have been applied. While some steps increase thermal noise (e.g., step 2, which removes entire baselines), they are effective in reducing RFI-related contamination. In all panels, the foreground horizon line is shown as a solid red line, while the green and black dashed lines indicate the delay ranges where most of the power from Cas\,A and Cygnus\,A is expected, respectively.}
    \label{fig:ps2d_post_cal_flag_steps}
\end{figure*}

Similar to \citetalias{Mertens20}, we use \textsc{Sagecal-CO} with high-regularisation parameters optimised to minimise the gain variations per individual cluster. The solution time interval varies between 2.5 and 20 min depending on the total apparent flux in a cluster. To ensure maximum smoothness, fitted smooth third-order Bernstein polynomials are used instead of the regularized gains, which may still contain higher-frequency spectral features~\footnote{These higher-frequency spectral features are unphysical and therefore could overfit the data.}. The sky-model multiplied with these smooth gains is then subtracted from the visibilities for further processing.

\subsection{Post-calibration Flagging}
\label{sec:post_cal_flagging}

In the \citetalias{Mertens20} analysis, we also encountered a substantial amount of low-level RFI, which we believe significantly contributed to the excess power observed on small baselines. Some measures were implemented in the previous analysis to mitigate this impact. Notably, we systematically flagged certain small baselines that were visibly heavily affected by RFI. Subsequent near-field imaging~\citep[see][for the methodology]{Smeenk20} confirmed that the source of this interference was local to the superterp\footnote{The Superterp consists of 24 densely packed stations within a 300-meter diameter area at the core of LOFAR.}. In the current analysis, we aimed to improve upon this approach by automating the detection and flagging of low-level RFI. Our post-processing flagging procedure includes the following steps:

\begin{description}
    \setlength\itemsep{1em}

    \item[Wide bandwidth AOFlagger:] The AOFlagger algorithm~\citep{Offringa12} was applied, post point-source subtraction, across the full frequency band of each of the three redshift bins. This increases our sensitivity to fainter and broadband RFI. The effect of this step is shown in the second panel of \autoref{fig:ps2d_post_cal_flag_steps}, where RFI contamination along the horizon line and within the foreground wedge is visibly reduced.  

    \item[Timeslots flagging:] Timeslots with more than 35\% of flagged data and baselines with over 80\% of flagged data are fully flagged, with the aim to minimise the impact of flagging caused by missing frequency channels. The issue with missing frequency channels arises because flagged samples introduce spectral discontinuities, causing excess power when the data are averaged or gridded, which can bias the 21-cm power spectrum~\citep{Offringa19a}.  

    \item[Flag RFI in the 147--149 and 155--158 MHz bands:] By visual inspection, we noticed that some baselines were affected by broadband RFI in these bands. The detection and flagging of this RFI was automated using a statistical thresholding based on the mean power ratio inside and outside the band. The number of flagged baselines for this step is shown in the shaded and blank bars in \autoref{fig:vis_flagger}. Typically, only a few baselines were affected by this RFI. The 147--149 MHz band RFI mostly affects earlier observations (cycles 0 to 2), while the 155--158 MHz band RFI primarily affects cycles 2 and 3, impacting not only small baselines but also longer ones.  
    An example of the baselines affected by this type of RFI is presented in~\hyperref[sec:ap_rfi_baselines]{Appendix~\ref*{sec:ap_rfi_baselines}}.  

    \item[Delay-space baseline flagger:] Finally, a systematic search for RFI-corrupted baselines was conducted. This involved computing the delay spectra for each baseline individually, analysing the median amplitude over time above the horizon limit, and flagging baselines that showed outlier peaks beyond a set threshold.  
    The effect of this step is shown in the third panel of \autoref{fig:ps2d_post_cal_flag_steps}. The number of flagged baselines for this step is represented by the plain bars in \autoref{fig:vis_flagger}. Examples of baselines flagged by this method are presented in~\hyperref[sec:ap_rfi_baselines]{Appendix~\ref*{sec:ap_rfi_baselines}}. Although it results in a non-negligible increase in thermal noise on these baselines due to fully flagging entire baselines, the benefit of reducing these contaminants far outweighs the cost. For future work, we are investigating the possibility of filtering these contaminants—mainly from local RFI sources—to preserve as much data as possible.
\end{description}

The cumulative effect of these flagging steps, in addition to other flagging steps on the gridded visibility cubes described later, is shown in \autoref{fig:ps2d_post_cal_flag_steps}.

\subsection{Imaging}

\begin{figure}
    \includegraphics{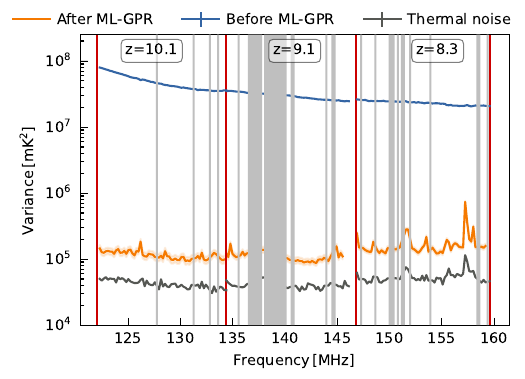}
    \caption{Variance as function of frequency for the 10 combined nights of the three redshift bins. The top line (blue) shows the Stokes-I power after sky-model subtraction (DD-calibration). The middle line (orange) shows the variance of the residual after ML-GPR, the bottom line (dark grey) show the thermal noise level estimated from the gridded time-differenced visibilities. Flagged channels are shown in light grey.}
    \label{fig:variance_combined_nights}
\end{figure}

\begin{figure}
    \includegraphics{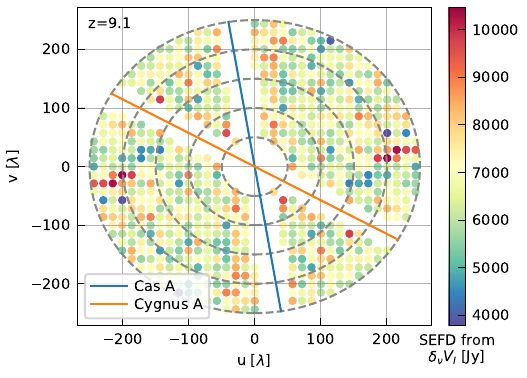}
    \caption{Noise standard deviation, in SEFD units, estimated from the channel-differenced noise at each $uv$-cell, after sky-model subtraction. Data are shown for the Stokes-I 10 nights combined $z \approx 9.1$ data cube. The trace of the Cas\,A and Cygnus\,A sources in the $uv$- plane, which are fully flagged, are denoted with a blue and orange line respectively.}
    \label{fig:uv_grid_flagging}
\end{figure}

After calibration, source subtraction, and RFI excision, the residual visibilities are gridded and imaged independently for each sub-band using \textsc{WSClean}\footnote{https://sourceforge.net/projects/wsclean/}~\citep{Offringa14}, producing an $(l, m, f)$ image cube, with $f$ the frequency. Separate Stokes-I and V images (in Jy PSF$^{-1}$), as well as PSF (point spread function) maps, are generated for each sub-band using natural weighting. Additionally, even and odd 10-second time-step images are created, enabling the generation of gridded time-differenced visibilities to estimate the thermal noise variance in the data. The sub-bands are then combined into image cubes with an 8.5$\degree\times$8.5$\degree$ field of view and a pixel size of 0.5 arcmin. These cubes are trimmed using a Tukey spatial filter with a $4\degree$ diameter, focusing on the primary beam’s most sensitive region (the station's primary beam full width at half-maximum (FWHM) at 140 MHz is approximately 4.1$\degree$). The Tukey window is chosen to find a compromise between avoiding sharp image edges and maximizing the observed volume, thereby improving sensitivity.

The image cubes produced by \textsc{WSClean} are converted to Kelvin units following the procedure outlined in~\citetalias{Mertens20}, and then transformed back to gridded visibilities through an inverse Fourier transform applied independently for each frequency channel. For each data set, the gridded visibilities $V(u, v, f)$ are stored in HDF5 format in Kelvin units, along with the number of visibilities that contribute to each $(u, v, f)$ grid point, $N_{\mathrm{vis}}(u, v, f)$.

A final outlier rejection is performed on the gridded visibility cubes to flag any remaining low-level RFI using a simple threshold-clipping method. Channels are flagged based on outliers in Stokes-V and Stokes-I variance. The flagged frequencies for the three redshift bins are shown in~\autoref{fig:variance_combined_nights} (grey areas). Many channels are flagged in the $z\approx9.1$ redshift bin, while the $z\approx10.1$  bin is comparatively cleaner. $uv$-cells are also flagged based on outliers in the weights, Stokes-V variance, and channel-difference Stokes-I variance. Additionally, $uv$-cells corresponding to sidelobes originating from Cas\,A and Cygnus\,A in $uv$-space are flagged~\citep[see][for a formal mathematical treatment of this effect]{Munshi25}. The flagged $uv$-cells for the three redshift bins are shown in~\autoref{fig:uv_grid_flagging}, and most flagging is due to Cas\,A and Cygnus\,A. The impact of this flagging on the cylindrically averaged power spectra is shown in~\autoref{fig:ps2d_post_cal_flag_steps}, which indicates that the A-team flagger primarily reduces power in regions where these sources are expected to contribute, helping to reduce contaminants.

\subsection{Combining nights}

To achieve better sensitivity and reduce thermal noise (and other incoherent errors), combining data from multiple nights of observation is essential. In this analysis, data from 14 nights are processed, but only the best 10 nights for each redshift bin are selected for combination. This allow us to be consistent with the~\citetalias{Mertens20}, while also allowing for the exclusion of clearly suboptimal nights. The selected nights are listed in~\autoref{table:observations}. This selection is based on the quality of each nights, as reflected in the post-calibration cylindrically averaged power spectra (see~\hyperref[sec:ap_all_ps2d]{Appendix~\ref*{sec:ap_all_ps2d}} for a complete list of cylindrically averaged power spectra for all analysed nights).

The selected nights are combined at the visibility level, following the weighting scheme:
\[
V_{\mathrm{c}n}(u, v, f) = \frac{\sum_{i=1}^{n} V_i(u, v, f) W_i(u, v, f)}{\sum_{i=1}^{n} W_i(u, v, f)}
\]
where $V_i$ represents the visibility cube for the $i$-th night, $V_{\mathrm{c}n}$ is the combined visibility cube for $n$ nights, and $W_i$ is the weights cube for the $i$-th night indicating the effective number of visibilities contributing to each $uv$-grid point.

\subsection{Residual foregrounds removal}
\label{sec:ml_gpr}

\begin{figure}
    \includegraphics{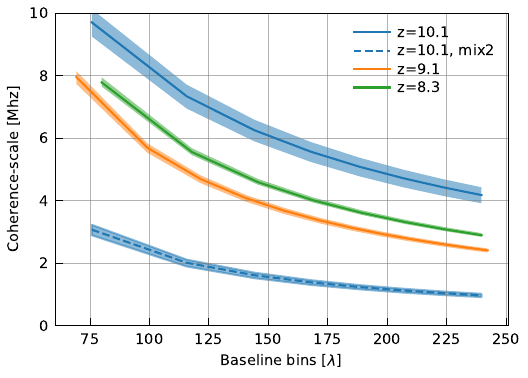}
    \caption{Coherence scales of the mode-mixing foreground GPR components for the three different redshift bins. The baseline dependence of the coherence scale is defined by~\autoref{eq:fg_baseline_dependence}. Based on the estimated values of the $\theta$ parameters for the mode-mixing components, the coherence scales range approximately between $8\,\mathrm{MHz}$ and $\approx2.5\,\mathrm{MHz}$ for redshift bins 8.3 (green line) and 9.1 (orange line). For redshift bin 10.1 (blue line), two mode-mixing components are required, yielding coherence scales ranging approximately between $10\,\mathrm{MHz}$ and $4\,\mathrm{MHz}$ for the first component (solid line), and between $3\,\mathrm{MHz}$ and $1\,\mathrm{MHz}$ for the second (dashed line).}
    \label{fig:gpr_model_fg_coherence_scale}
\end{figure}

After calibration and sky-model source subtraction, the residual Stokes-I visibilities, \( V_{\text{res}} \), is still dominated by foregrounds, predominantly extra-galactic emission below the confusion limit and partially-polarized diffuse Galactic emission. These foregrounds are approximately three orders of magnitude brighter than the 21-cm signal. Following the methodology of \citetalias{Mertens20}, we apply Gaussian Process Regression~\citep[GPR,][]{Rasmussen06} to model and subtract the residual foregrounds. This approach exploits the distinct frequency coherence scales of the foregrounds and the 21-cm signal, and follow the methodology established in~\cite{Mertens18a}. The data \( \mathbf{d} \) are modelled as:
\begin{equation}
\mathbf{d}(f) = \mathbf{f}_{\text{fg}}(f) + \mathbf{f}_{21}(f) + \mathbf{n}(f),
\end{equation}
where the vectors $\mathbf{f}_{\text{fg}}$, $\mathbf{f}_{21}$, and $\mathbf{n}$ represent the foregrounds, 21-cm signal, and noise components, respectively, as functions of frequency $f$. The different components, that we assume independent, can be separated based on their distinct spectral behaviour~\citep{Mertens18a}\footnote{Currently, our GP model does not include spatial (in the plane of the sky) correlation, this is left for a future development.}. This behaviour is defined by the covariance of the components between frequency channels. The total frequency-frequency covariance matrix of the data, $\mathbf{K}$, is a sum of the individual covariance matrices, encompassing the foregrounds covariance matrix $\mathbf{K}_{\text{fg}}$, the 21-cm covariance matrix $\mathbf{K}_{21}$, and the noise covariance matrix $\mathbf{K}_n$:
\begin{equation}
\mathbf{K} = \mathbf{K}_{\text{fg}} + \mathbf{K}_{21} + \mathbf{K}_n,
\end{equation}
using the shorthand $\mathbf{K} \equiv \mathbf{K}(f, f)$. The joint probability density distribution of the observations $\mathbf{d}$ and the function values $ \mathbf{f}_{\mathrm{fg}}$ of the foreground model at the same frequencies $f$ are then given by,
\begin{equation}
\begin{bmatrix} \mathbf{d} \\ \mathbf{f}_{\mathrm{fg}} \end{bmatrix} \sim  
\mathcal{N}\left( \begin{bmatrix} \mathbf{0} \\ \mathbf{0} \end{bmatrix}, \begin{bmatrix} 
\mathbf{K}_{\mathrm{fg}} + \mathbf{K}_{21} + \mathbf{K}_{\mathrm{n}} & \mathbf{K}_{\mathrm{fg}} \\ 
\mathbf{K}_{\mathrm{fg}} & \mathbf{K}_{\mathrm{fg}} \end{bmatrix} \right).
\end{equation}The foreground model is then a Gaussian process, conditional on the data:
\begin{equation}
\label{eq:gpr_fg_fit_distribution}
\mathbf{f}_{\mathrm{fg}} \sim \mathcal{N}\left({\cal E}(\mathbf{f}_{\mathrm{fg}}), \mathrm{cov}(\mathbf{f}_{\mathrm{fg}})\right),
\end{equation}
with expectation value and covariance defined by:
\begin{align}
\label{eq:gpr_predictive_mean_eor}
\mathcal{E}(\mathbf{f}_{\mathrm{fg}}) &= \mathbf{K}_{\mathrm{fg}}\mathbf{K}^{-1} 
\mathbf{d} \\
\label{eq:gpr_predictive_cov_eor}
\mathrm{cov}(\mathbf{f}_{\mathrm{fg}}) &= \mathbf{K}_{\mathrm{fg}} - \mathbf{K}_{
\mathrm{fg}}\mathbf{K}^{-1}\mathbf{K}_{\mathrm{fg}}.
\end{align}
The residual is obtained by subtracting ${\cal E}(\mathbf{f}_{\mathrm{fg}})$ from the observed data:
\begin{equation}
\mathbf{r} =\mathbf{d} - {\cal E}(\mathbf{f}_{\mathrm{fg}}).
\end{equation}
The GPR method utilises parametrized covariance functions to model the different components of the signal. These functions are defined by a set of parameters, $\boldsymbol{\theta}$, that control properties like variance and the overall shape of the covariance function. In GPR, we perform prior covariance model selection under a Bayesian framework by choosing the model that maximizes the log-marginal-likelihood.
\begin{equation}
\label{eq:hyper_lml}
\log p(\mathbf{d} \mid f, \boldsymbol{\theta}) = -\frac{1}{2}
\mathbf{d}^\intercal \mathbf{K}^{-1} \mathbf{d}
- \frac{1}{2} \log |\mathbf{K}| - \frac{n}{2} \log 2\pi,
\end{equation}
with $n$ the number of data points. The posterior probability density of the parameters is then estimated by applying Bayes' theorem, incorporating the prior on the parameters:
\begin{equation}
\label{eq:hyper_post}
\log p(\boldsymbol{\theta} \mid \mathbf{d}, f) \propto \log p(\mathbf{d} \mid f, \boldsymbol{\theta}) + \log p(\boldsymbol{\theta}).
\end{equation}
The posterior distributions for the parameters of our GP model are derived for each redshift bin with a nested sampling algorithm using the \textsc{UltraNest}\footnote{\url{https://johannesbuchner.github.io/UltraNest/}} package~\citep{Buchner21}.

To enhance the robustness of our methodology, we have introduced several improvements over the ~\citetalias{Mertens20} approach. One key limitation of the earlier approach was the reliance on an exponential covariance function to model the 21-cm signal. While this choice was effective for a subset of simulations, it risks introducing biases when applied to a wider range of scenarios~\citep{Kern21}. In this work, we address this limitation by employing a machine learning approach to construct a physically motivated covariance function directly from simulations of the 21-cm with different astrophysics. Specifically, we use a Variational Autoencoder (VAE) to learn a low-dimensional representation of the 21-cm signal covariance from a large ensemble of simulations~\citep{Mertens24}. This enables a more accurate and flexible modelling of the 21-cm component in the data.

\begin{table*}
\centering
\caption{Summary of the Gaussian Process (GP) model, listing each component's covariance function, the priors on its parameters, and the estimated medians with 68\% confidence intervals obtained using a nested sampling procedure. All variance parameters are expressed relative to the noise variance estimated from the time-differenced data. The prior ranges for the variance parameters of the different components are chosen to be very broad, having no impact on this analysis. The 21-cm component is learned from \textsc{Grizzly} simulations. The coherence-scale of the mode mixing components is baseline dependent, as defined by the foregrounds wedge equation. The corresponding coherence-scales for each redshift bins can be seen in~\autoref{fig:gpr_model_fg_coherence_scale}.}
\begin{threeparttable}
\label{tab:gp_model_fit}
\renewcommand{\arraystretch}{1.5} % Increase row height
\begin{tabularx}{\textwidth}{l l| *{6}{>{\centering\arraybackslash}X}}
\toprule
\multirow{2}{*}{\makecell[l]{\textbf{Component}\\Covariance function$^a$}} & \multirow{2}{*}{Parameter} & \multicolumn{2}{c}{$z\approx8.3$} & \multicolumn{2}{c}{$z\approx9.1$} & \multicolumn{2}{c}{$z\approx10.1$} \\
\cmidrule(lr){3-4} \cmidrule(lr){5-6} \cmidrule(lr){7-8} 
& &  Prior & Posterior & Prior & Posterior & Prior & Posterior \\
\toprule
\multirow{2}{*}{\makecell[l]{\textbf{Sky}\\RBF}} & $\sigma^2$ & $-$ & $393.5\substack{+16.4\\-16.4}$ & $-$ & $680.3\substack{+28.6\\-27.9}$ & $-$ & $1059.5\substack{+54.4\\-52.1}$ \\
& $l$ [MHz] & $=80$ & $=80$ & $\mathcal{U}(30, 60)$ & $41.8\substack{+6.5\\-4.8}$ & $\mathcal{U}(10, 50)$ & $31.8\substack{+5.2\\-3.4}$ \\
\midrule
\multirow{3}{*}{\makecell[l]{\textbf{Mode mixing}\\Matérn 3/2}} & $\sigma^2$ & $-$ & $81.5\substack{+2.7\\-2.7}$ & $-$ & $114.8\substack{+4.6\\-4.4}$ & $-$ & $290.5\substack{+25.7\\-23.5}$ \\
& $\theta$ [radians] & $\mathcal{U}(0.15, 0.25)$ & $0.210\substack{+0.004\\-0.004}$ & $\mathcal{U}(0.2, 0.3)$ & $0.238\substack{+0.006\\-0.006}$ & $\mathcal{U}(0.05, 0.15)$ & $0.106\substack{+0.008\\-0.008}$ \\
& $\delta_{\text{buffer}}$ [\micro s] & $=0.02$ & $=0.02$ & $=0.01$ & $=0.01$ & $=0.04$ & $=0.04$ \\
\midrule
\multirow{2}{*}{\makecell[l]{\textbf{Mode mixing 2}\\Matérn 3/2}} & $\sigma^2$ & $-$ & $-$ & $-$ & $-$ & $-$ & $18.8\substack{+3.0\\-2.9}$ \\
& $\theta$ [radians] & $-$ & $-$ & $-$ & $-$ & $\mathcal{U}(0.4, 0.8)$ & $0.57\substack{+0.04\\-0.04}$ \\
\midrule
\multirow{2}{*}{\makecell[l]{\textbf{Excess}\\RBF}} & $\sigma^2$ & $-$ & $1.08\substack{+0.06\\-0.06}$ & $-$ & $1.4\substack{+0.1\\-0.1}$ & $-$ & $1.3\substack{+0.2\\-0.1}$ \\
& $l$ [MHz] & $\mathcal{U}(0.3, 0.6)$ & $0.41\substack{+0.01\\-0.01}$ & $\mathcal{U}(0.2, 0.3)$ & $0.40\substack{+0.02\\-0.02}$ & $\mathcal{U}(0.4, 0.8)$ & $0.36\substack{+0.02\\-0.02}$ \\
\midrule
\multirow{3}{*}{\makecell[l]{\textbf{21-cm}\\Learned kernel}} & $\sigma^2$ & $-$ & $< 0.02$ & $-$ & $< 0.05$ & $-$ & $< 0.02$ \\
& $x_1$ & $\mathcal{U}(-3, 3)$ & $0.0\substack{+2.0\\-1.9}$ & $\mathcal{U}(-3, 3)$ & $-0.2\substack{+2.1\\-1.9}$ & $\mathcal{U}(-3, 3)$ & $0.2\substack{+1.9\\-2.1}$ \\
& $x_2$ & $\mathcal{U}(-3, 3)$ & $0.9\substack{+1.5\\-2.3}$ & $\mathcal{U}(-3, 3)$ & $-0.6\substack{+2.2\\-1.6}$ & $\mathcal{U}(-3, 3)$ & $-0.1\substack{+2.2\\-1.9}$ \\
\midrule
\makecell[l]{\textbf{Noise}\\Identity} & $\alpha$ & $\mathcal{U}(1.2, 1.9)$ & $1.74\substack{+0.01\\-0.01}$ & $\mathcal{U}(1.4, 1.7)$ & $1.59\substack{+0.02\\-0.02}$ & $\mathcal{U}(1.1, 1.5)$ & $1.31\substack{+0.02\\-0.02}$ \\
\bottomrule
\end{tabularx}
\begin{tablenotes}
  \small
  \item[a] `RBF' is Radial Basis Function; `Matérn 3/2' and `Matérn 5/2' are Matérn class functions~\citep{Rasmussen06} with $\nu=3/2$ and $\nu=5/2$ respectively.
\end{tablenotes}
\end{threeparttable}
\end{table*}

The simulations used for this training are based on the \textsc{Grizzly} framework~\citep{Ghara15,Ghara18}, which encompasses a wide range of astrophysical scenarios. The trained VAE kernel thus serves as an effective model for the 21-cm signal covariance matrix, parametrized by three key factors: two latent space dimensions, \(x_1\) and \(x_2\), and a variance scaling factor, \(\sigma^2_{21}\). This machine-learning-derived covariance function is integrated into our GPR framework to more accurately isolate the 21-cm signal from the foregrounds and systematics. We refer to~\cite{Mertens24} for a comprehensive description of the VAE method, and~\cite{Acharya24} for the more specific description of the \textsc{Grizzly} trained VAE kernel.

For the foreground components, we use analytical covariance models from the Matérn class~\citep{Rasmussen06}, which provide a flexible description of the spectral behaviour observed in the data. Different forms within the Matérn class, such as Matérn 3/2 and Matérn 5/2, have been tested through multiple trials to identify the most suitable model for our observations.

Additionally, to account for the baseline dependence of the foregrounds coherence scale (which manifests as the `wedge' in \(k\)-space), we set the coherence scale as a function of baseline length using:
\begin{equation}
\label{eq:fg_baseline_dependence}
l_{\text{mix}}(u) = \frac{f_m}{f_m \delta_{\text{buffer}} + b \sin \theta},
\end{equation}
where \(l_{\text{mix}}\) is the mode-mixing coherence-scale, \(\delta_{\text{buffer}}\) is the delay buffer, \(b\) is the baseline length, \(\theta\) is the angular distance of foreground sources from the phase center, and \(f_m\) is the central frequency of the redshift bin.

The same range of covariance models has also been successfully used in the recent SKA Data Challenge 3a~\citep{bonaldi25}, where they were tested on externally produced data and emerged as the top-performing approach. While this demonstrates the performance of the method, it should be noted that the challenge dataset was relatively simplified compared to real observational data.

Our findings indicate that the data cannot be adequately described by only the foreground and 21-cm components. There is additional power within the data characterized by a smaller coherence scale (typically of $\approx0.4$ MHz), making it more challenging to differentiate from the 21-cm signal. This necessitates the addition of an `Excess' component to our covariance model to ensure a more effective separation between the foregrounds and the 21-cm signal. 

The final parametric GP model is composed of five terms:
\begin{equation}
  \rm K = K_{\mathrm{sky}} + K_{\mathrm{mix}}  + K_{\mathrm{21}} + \alpha K_{\mathrm{n}} + K_{\mathrm{ex}}.
\end{equation}
where $K_{\mathrm{n}}$ represents the thermal noise diagonal covariance matrix, estimated from the time-differenced visibility cube, and $\alpha$ a scaling factor that accounts for additional frequency-uncorrelated noise in the data in excess of the thermal noise.

The various components and their parameters are detailed in~\autoref{tab:gp_model_fit}. The model includes six main components:
\begin{enumerate}[leftmargin=1.5em,label=(\roman*)]
% \begin{description}
    \setlength\itemsep{0.5em}
    \setlength\labelwidth{0pt}
    % \setlength\itemsep{0.25em}
    % \item[(1) The foregrounds ---]
    \item \textbf{The sky ---} Represents the large-scale spectrally-smooth foregrounds, the astrophysical sources, Galactic and extra-galactic, in the field of view of the image cube.
    \item \textbf{The Mode-mixing 1 ---} Accounts for chromatic effects with shorter coherence-scale introduced by the spectral response of the instrument.
    \item \textbf{The Mode-mixing 2 ---} An additional mode-mixing component used exclusively for the $z \approx 10.1$ bin, as one component was insufficient to account for the chromatic effects for this redshift.
    \item \textbf{The excess ---} Captures small-scale residual structures not fully accounted for by the foreground models.
    \item \textbf{The 21-cm ---} The learned component that should capture the 21-cm signal in the data.
    \item \textbf{The noise ---} Represents the thermal noise inherent to the observation.
\end{enumerate}

Each of these components has different associated parameters, for which we assigned prior ranges that were iteratively refined during model development. The prior ranges and the posterior values reflect our best understanding of the spectral behaviours of the foregrounds and excess components for each redshift bin, and ensure the separability between the 21-cm signal in one part and the foregrounds and instrumental effects in another part. In addition to selecting the model and prior ranges that would maximize the Bayesian evidence, this process is also guided by injection tests. It consists of injecting a mock 21-cm signal into the data and ensuring that it can be recovered by our component separation process (this procedure is fully described in~\autoref{sec:validations}).

\subsection{Power spectra estimation}
\label{sec:power_spectra_definition}

We define the cylindrically averaged power spectrum as~\citep{Mertens20}:
\begin{equation}
P(k_{\perp}, k_{\parallel}) = \frac{X^2 Y}{\Omega_{\mathrm{PB}} B} \left<\left|
\tilde{V}
(u, v, \tau)\right|^2\right>,
\end{equation}
where $\tilde{V}(u, v, \tau)$ is the Fourier transform of the visibility cube $V(u, v, f)$ in the frequency direction, $B$ is the frequency bandwidth, $\Omega_{\mathrm{PB}}$ is the integral of the square of the primary beam gain over solid angle, X and Y are conversion factors from angle and frequency to comoving distance, and $<..>$ denotes the averaging over baselines inside a bin-width. The Fourier modes are in units of inverse comoving distance and are given by~\citep{Morales06,Trott12}:
% \begin{align}
% k_{\perp} = \frac{2 \pi |\mathbfit{u}|}{D_M(z)},\ 
% k_{\parallel} = \frac{2 \pi H_0 f_{21} E(z)}{c(1+z)^2} \tau,\ 
% k = \sqrt{k_{\perp}^2 + k_{\parallel}^2},
% \end{align}
where $D_M(z)$ is the transverse co-moving distance, $H_0$ is the Hubble constant, $f_ {21}$ is the frequency of the hyperfine transition, and $E(z)$ is the dimensionless Hubble parameter. We also define the dimensionless power spectrum by averaging the power spectrum in spherical shells as:
\begin{equation}
\Delta^2({k}) = \frac{k^3}{2 \pi^2} P(k).
\end{equation}
This representation is well suited for characterising the 21-cm signal, to a first order\footnote{Although the 21-cm power spectrum is expected to present slight anisotropies due to redshift space distortions and light cone effects, it is common practice to ignore these to the first order and use the spherically-averaged power spectrum.}. We limit our analysis to a bandwidth of 12 MHz to limit the effects of signal evolution, i.e. the light-cone effect~\citep{Datta12}.

When displaying the cylindrically averaged power spectra, we overplot the horizon line. This line represents the limit above which we do not expect any foreground emission and delimits the foreground wedge. The line is computed using improved calculations that take into account the sky curvature and phase-referencing away from zenith \citep{Munshi25}.

\subsubsection{Uncertainty calculation}

To compute the power spectrum and its uncertainty for a given ML-GPR component, such as the 21-cm signal for illustration as given below, we employ the following procedure:
\begin{enumerate}[leftmargin=*,label=(\roman*)]
    \item We draw $m$ samples from the posterior distribution of the parameters obtained from the nested sampling results from the GPR analysis.
    
    \item For each set of parameters, we calculate the predictive mean $\mathcal{E}(\mathbf{f}_{21})$ and the predictive covariance $\mathrm{cov}(\mathbf{f}_{21})$ of the 21-cm component.
    
    \item For each parameter sample, we generate a realisation of the 21-cm signal by adding a random fluctuation to the predictive mean:
    \begin{equation}
    \mathbf{f}_{21} = \mathcal{E}(\mathbf{f}_{21}) + \delta_{21},
    \end{equation}
    where $\delta_{21}$ is a random vector drawn from a multivariate Gaussian distribution with covariance $\mathrm{cov}(\mathbf{f}_{21})$.
    
    \item We compute the power spectrum for each realisation $\mathbf{f}_{21}$ using the definitions provided earlier.
    
    \item After processing all $m$ samples, we have $m$ power spectra. We estimate the final power spectrum of the 21-cm component by calculating the median of these $m$ spectra at each $k$-value. The associated 1$\sigma$ uncertainty is determined by the standard deviation of the power spectra values at each $k$.
\end{enumerate}

This method accounts for uncertainties in the parameters and propagates them through to the power spectrum estimation. By incorporating the variability from both the parameters and the intrinsic fluctuations of the 21-cm signal, we obtain a more robust estimate of the power spectrum and its statistical uncertainty. Note that calibration gain errors are ignored in this calculation. Calibration errors manifest themselves as additional power ({\emph solver noise} in the spectra when the solutions are transferred from longer to shorter baselines. \cite{Mevius22} have shown that the expected level of solver noise power is well below that of thermal noise.

\subsubsection{ML-GPR inpainting for Stokes-I and foreground power spectra estimation}

When estimating power spectra from Stokes-I data (after DD-calibration) or from the foreground components, missing frequency channels -- flagged by our post-calibration procedure (see \autoref{fig:variance_combined_nights}) -- can introduce strong spectral leakage during the delay transform. This leakage occurs because frequency gaps disrupt the continuity required for accurate Fourier transforms, leading to artefacts that contaminate the power spectrum estimation.

To mitigate this issue for Stokes-I data and foreground components, we use the results from ML-GPR to interpolate the data values at the missing channels. Importantly, this interpolation method is only applied to these components and not to others, such as the residuals, the 21-cm signal, or the excess component; thus, our final results remain unaffected by this method.

By leveraging the covariance structure of the data, ML-GPR allows us to obtain estimates of the data at the flagged frequencies. Given the unflagged frequency channels $f$ and the flagged channels $f_*$, we compute the predictive mean at the flagged channels using:
\begin{equation}
\label{eq:gpr_predictive_mean_unobserved}
\mathcal{E}(\mathbf{d}_*) = \mathbf{K}(f_*, f)\, \mathbf{K}(f, f)^{-1}\, \mathbf{d}.
\end{equation}
Similarly, we compute the predictive mean of the foreground component over the full frequency range $f_{\mathrm{full}}$, including the flagged channels, using:
\begin{equation}
\label{eq:gpr_predictive_mean_full}
\mathcal{E}(\mathbf{f}_{\mathrm{fg, full}}) = \mathbf{K}_{\mathrm{fg}}(f_{\mathrm{full}}, f)\, \mathbf{K}(f, f)^{-1}\, \mathbf{d}.
\end{equation}
By interpolating across the missing channels, we reconstruct a continuous frequency spectrum, reducing spectral leakage effects in the delay transform. This 'inpainting' method, similar to that described by \citet{Kern21}, is used only to estimate the power spectra of the Stokes-I data and the foreground components, and is not applied to other components or the final results.

\begin{figure*}
    \includegraphics{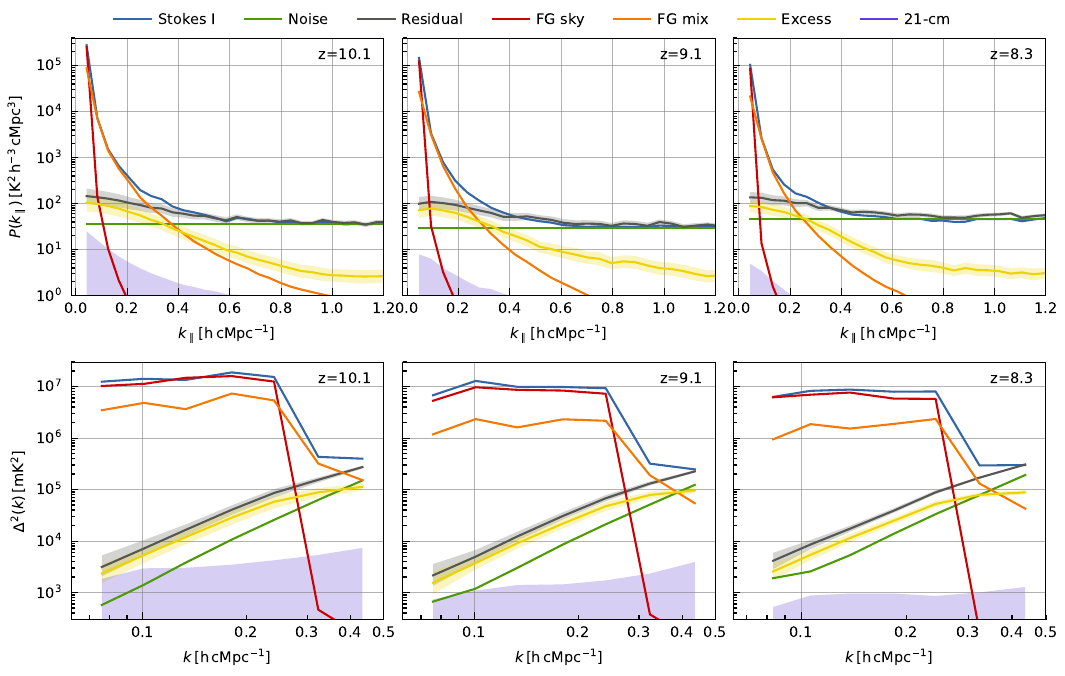}
    \caption{ML-GPR components decomposition of the residual Stokes-I for the three redshift bins. The residual Stokes-I data (blue line) are decomposed into the following components: foreground sky (red), mode-mixing (orange), excess (yellow), noise (green), and the 21-cm signal (magenta). The residual (grey line) represents the Stokes-I data after subtracting all the foreground components. The shaded area represents the 2-$\sigma$ uncertainty. The top panel displays the cylindrically-averaged power spectra as a function of $k_{\parallel}$ (averaged over $k_{\perp}$), while the bottom panels shows the spherically-averaged power spectra.}
    \label{fig:ps_gpr_decomposition}
\end{figure*}

\begin{figure*}
    \includegraphics{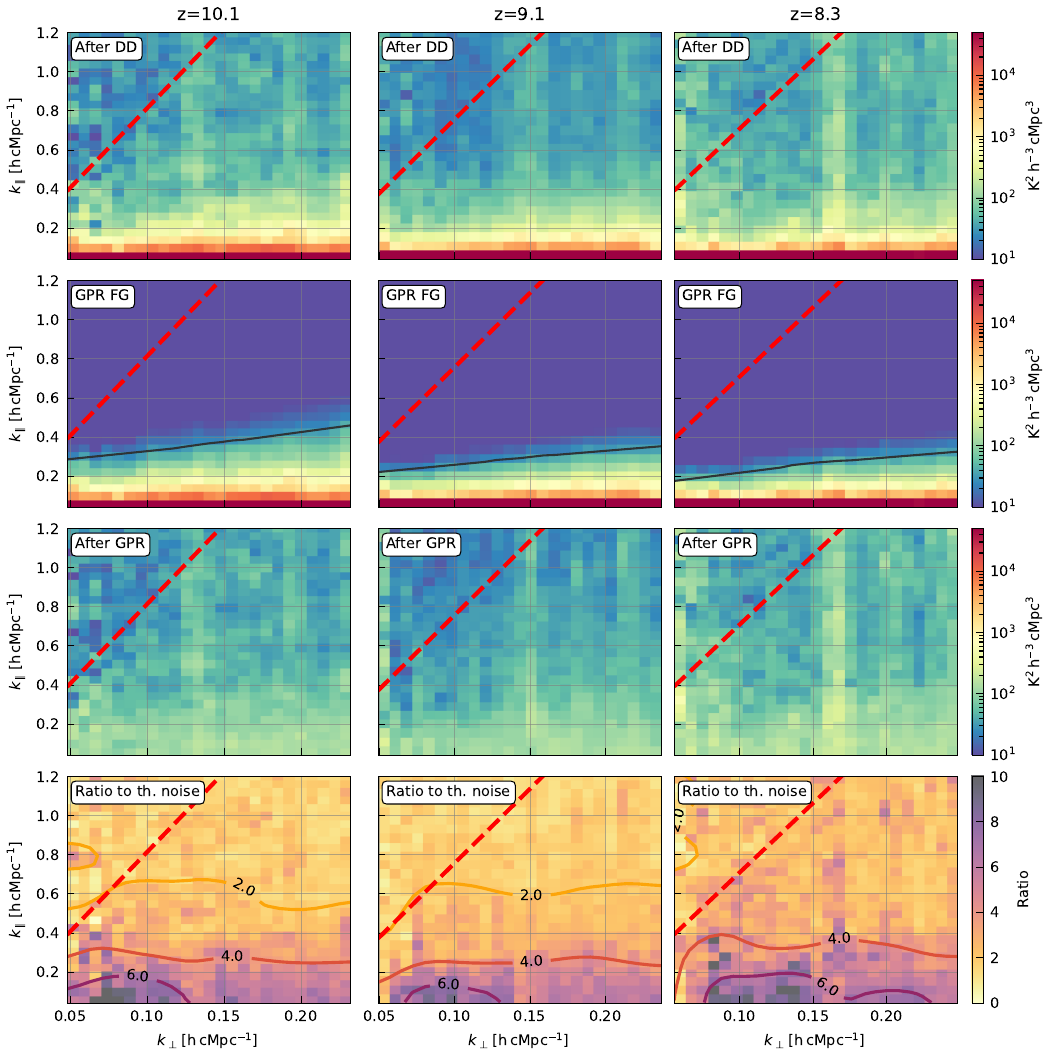}
    \caption{Cylindrically averaged power spectra of the ML-GPR component decomposition of the residual Stokes-I data for the three redshift bins. The top row shows the power spectra before ML-GPR. The second row shows the power spectra of the foreground components only (comprising the `sky' and `mode-mixing' components). The third row presents the power spectra of the residual data after subtracting the foregrounds using ML-GPR. The last row shows the ratio of the residual power to the thermal noise power. In all panels, the foreground horizon line is depicted by a red dashed line. The black solid line in the second row delimits the regions below which the foregrounds power is higher than the thermal noise power.}
    \label{fig:ps2d_gpr_residual}
\end{figure*}

\section{Results}
\label{sec:results}

In this section, we present the results of our analysis, starting with the separation of components using ML-GPR. We then discuss the derived power spectra, compare our results with previous findings from~\citetalias{Mertens20}, and present new upper limits on the 21-cm signal power spectrum at redshifts $z\approx8.3$, $z\approx9.1$, and $z\approx10.1$.

\subsection{Component separation}
\label{sec:cmpt_sep}

\begin{figure*}
    \includegraphics{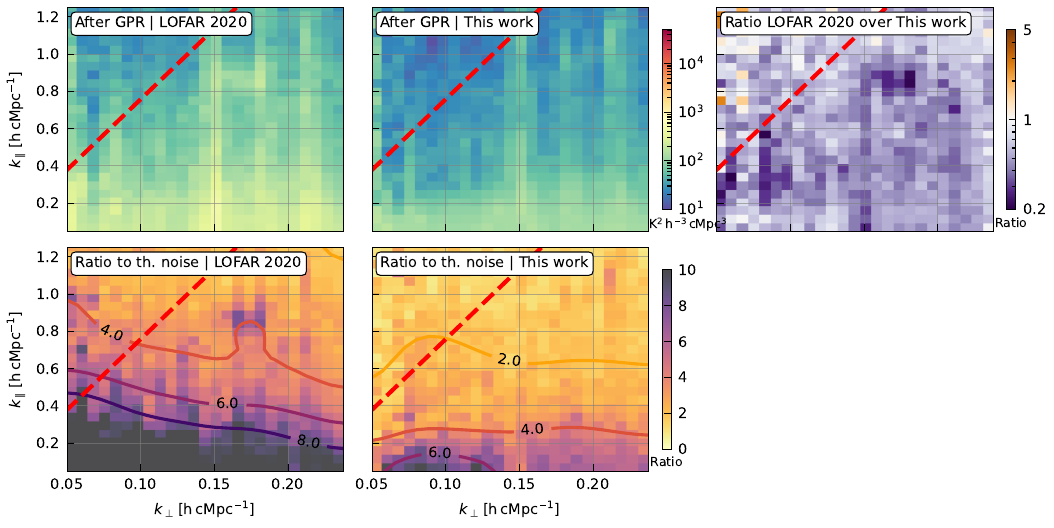}
    \caption{Comparison between the current results and those of \citetalias{Mertens20} at redshift $z \approx 9.1$. The top row shows the residual cylindrically averaged power spectra after GPR for the \citetalias{Mertens20} dataset (left panel), the current dataset (middle panel), and the ratio between the two (right panel). The bottom row presents the ratio of the residual power to the thermal noise for the \citetalias{Mertens20} dataset (left panel) and the current dataset (middle panel), accompanied by smoothed contour maps to emphasize the differences. In all panels, the foreground horizon line is indicated by a red dashed line. The significant reduction in residual power across the full $k$-space demonstrates the effectiveness of our improved data processing techniques.} 
    \label{fig:ps2d_2020_vs_2024}
\end{figure*}

We begin by analysing the results of the component separation. The priors and posterior estimates for the parameters of our ML-GPR model are presented in~\autoref{tab:gp_model_fit}, and detailed corner plots of the posterior distributions are provided in~\hyperref[sec:ap_all_corner]{Appendix~\ref*{sec:ap_all_corner}}. Most parameters are well constrained, except for the variance of the `21-cm' component for which, given the depth of the current data, only an upper limit is found. Moreover, we observe minimal correlation between the parameters, with the exception of a few intra-component parameters, such as between the variance and coherence scale of the mode-mixing component ($\sigma^2_{\mathrm{mix}}$ and $\theta_{\mathrm{mix}}$).

As expected, the `Sky' and `Mode-Mixing' components dominate the decomposition. The coherence scales of these components vary between redshift bins but align with theoretical expectations (see~\autoref{fig:gpr_model_fg_coherence_scale} for the coherence scales of the different mode-mixing components across the three redshift bins). Specifically, we observe a decrease in coherence scale as we decrease frequency (i.e., increase redshift). This trend is attributed to the increased intensity of the foreground emissions and the larger field of view of the station primary beam at lower frequencies. A larger field of view captures more emission from regions farther from the phase centre, which exhibit shorter coherence scales in frequency.

% For the $z \approx 10.1$ bin, we found that a single mode-mixing component was insufficient to account for the chromatic effects introduced by the instrument's spectral response. Consequently, we introduced a second mode-mixing component for this redshift bin to effectively capture these chromatic distortions.

The `Excess' component captures small-scale residual structures not fully accounted for by the foreground models. Overall, we observe that the coherence scale of the excess component is consistent across all redshift bins. Compared to the coherence scale observed in \citetalias{Mertens20}, where a coherence scale of 0.26\,MHz was found using a Matérn 5/2 kernel, we now find a coherence scale of approximately 0.4\,MHz using a Radial Basis Function (RBF) kernel. This indicates that the excess component in our current analysis is overall more frequency coherent. Possible sources of this residual excess are discussed in~\autoref{sec:discussion}.

The `21-cm' component remains largely unconstrained across all redshifts, and its variance is significantly lower than the noise variance. This is expected as we are using a relatively short total integration time. In this paper, we conservatively subtract only the foreground components (`sky' and `mode-mixing') from the data. We have not yet attempted to remove the `excess' component~\citep[see e.g.,][for a discussion on this]{Acharya24a}, but we may consider doing so in the future, once we are confident that it can be properly separated from the 21-cm signal.

\subsection{Power spectra}
\label{sec:result_ps}

% \begin{figure*}
%     \includegraphics{figures/ps2d_2020_vs_2024.pdf}
%     \caption{Comparison between current and~\citetalias{Mertens20} results. The top row shows the residual cylindrically-averaged power-spectra after GPR for the ~\citetalias{Mertens20} data set (left panel), current data set (middle panel) and the ratio between the two (right panel). The bottom rows show the shows the ratio of the residual power to the thermal noise power for ~\citetalias{Mertens20} data (left panel) and current data set (middle panel). In all panels, the foreground horizon line is depicted by a red dashed line. Smoothed contour maps is also shown for the bottom row.}
%     \label{fig:ps2d_2020_vs_2024}
% \end{figure*}

\autoref{fig:ps_gpr_decomposition} and~\autoref{fig:ps2d_gpr_residual} present the power spectra of the different components resulting from the ML-GPR decomposition. The foreground components are effectively described by a combination of a spectrally very smooth component and a spectrally less smooth component, both of which are well confined within the wedge region in $k$-space (see the second row of Figure~\ref{fig:ps_gpr_decomposition}). For small baselines ($k_{\perp}$ below $0.1\,h\,\mathrm{cMpc^{-1}}$), the foreground power even falls below the thermal noise level for $k_{\parallel} > 0.2\,h\,\mathrm{cMpc^{-1}}$ at redshifts $z \approx 8.3$ and $z \approx 9.1$, and for $k_{\parallel} > 0.3\,h\,\mathrm{cMpc^{-1}}$ at $z \approx 10.1$. This illustrates the more intense foreground power for the redshift bin $z \approx 10.1$.

The effective confinement of the foreground power within the wedge is crucial for the separability between the foregrounds and the 21-cm signal~\citep{Mertens18a}. Our improvements in calibration and RFI mitigation have significantly reduced spectrally correlated contaminants, leading to a better separability of the foregrounds from the 21-cm signal of interest.

The excess component has less spectral coherence than the foreground components. While it exceeds the noise power within the wedge regions, its influence decreases as \(k_{\parallel}\) increases. Even at high \(k_{\parallel}\), the excess remains only a small fraction of the noise power, indicating that while the excess component is not entirely negligible, it has a limited impact at higher \(k_{\parallel}\) modes, particularly above the foreground wedge.

The variance of the 21-cm signal component remains unconstrained across all three redshift bins. Consequently, the power spectra of this component are very low but with large uncertainties, reflecting the large uncertainties in the parameters associated with this component. This behaviour is consistent with the findings of~\citet{Mertens24}, where a scenario without a 21-cm signal was tested, resulting in similar uncertainties. The same pattern was also observed by~\cite{Acharya24} with LOFAR simulation when the 21-cm signal variance was much smaller than the noise variance. This consistency confirms that no 21-cm signal is detected in our data, or at least that it is not captured by the 21-cm component in our analysis.

After subtracting the foreground components from the data (after ML-GPR), the residual power spectra primarily consist of noise and the excess component, and are thus confined to low $k_{\parallel}$ modes. On average, the ratio of the residual power spectra to the thermal noise power spectra at high $k_{\parallel}$ (above $k_{\parallel} = 1\,h\,\mathrm{cMpc^{-1}}$) is approximately 1.6, 1.4 and 2.2 for the redshift bins $z \approx 10.1, 9.1, \text{ and } 8.3$ respectively. At lower $k_{\parallel}$ (below $k_{\parallel} = 0.2\,h\,\mathrm{cMpc^{-1}}$), the mean ratios are approximately 5.5, 5.0, and 5.9 for redshifts $z \approx 10.1, 9.1, \text{ and } 8.3$, respectively.

The observed ratios are lower for certain $k_{\perp}$ values where the noise is higher, like $k_{\perp} = 0.15\,h\,\mathrm{cMpc^{-1}}$ for example. This is because the ratio is computed relative to the thermal noise, and the excess component does not scale with the thermal noise. As a result, regions with higher noise exhibit lower ratios.

In a related note, the higher power observed in~\autoref{fig:ps2d_gpr_residual} (and in ~\autoref{sec:ap_all_ps2d}) before and after ML-GPR around $k_{\perp} \approx 0.15\,h\,\mathrm{cMpc^{-1}}$ is linked to the lower density of baselines of LOFAR in this range, when observing the NCP, which results in higher noise.

\subsection{Comparison with previous results}

Our updated analysis demonstrates a significant reduction in residual power across the entire $k$-range. \autoref{fig:ps2d_2020_vs_2024} presents a comparison between the cylindrically averaged power spectra from our \citetalias{Mertens20} analysis and the current analysis. On average, we observe a reduction in power inside the wedge by a factor of approximately two. In the \citetalias{Mertens20} analysis, the ratio of residual power to thermal noise was about 6.2 inside the wedge and approximately 3 above it. These ratios have been reduced to about 3 inside the wedge and 1.6 above it in our current analysis.

Notably, the most substantial reductions -- by factors of approximately 4 to 5 -- are observed at $k_{\perp} \approx 0.08$ and 0.16~$h\,\mathrm{cMpc}^{-1}$. These reductions are likely due to residual RFI, with the former being directly associated with local sources that we have effectively mitigated through delay-space baseline flagging techniques.

The application of delay-space baseline flagging (see \autoref{sec:post_cal_flagging}) has proven effective in reducing excess power on small baselines, which are particularly susceptible to RFI contamination. However, this improvement comes at the expense of an increase in thermal noise. Consequently, while the ratio of residual power to thermal noise has decreased significantly for small baselines, the ratio between the residuals from the \citetalias{Mertens20} and current analyses remains close to unity in this region.

These improvements are the cumulative result of enhancements implemented at every stage of our analysis: by refining the calibration procedures, improving sky model subtraction, optimising the GPR methodology, and enhancing RFI excision techniques, we have significantly reduced residual power levels using a dataset similar to that used in the~\citetalias{Mertens20} analysis. These methodological advancements have collectively contributed to bringing our measurements closer to the theoretical thermal noise limit. This demonstrates that careful attention to data processing details can lead to substantial improvements, even when using comparable observational data.

\begin{figure*}
    \centering
    \includegraphics[width=\textwidth]{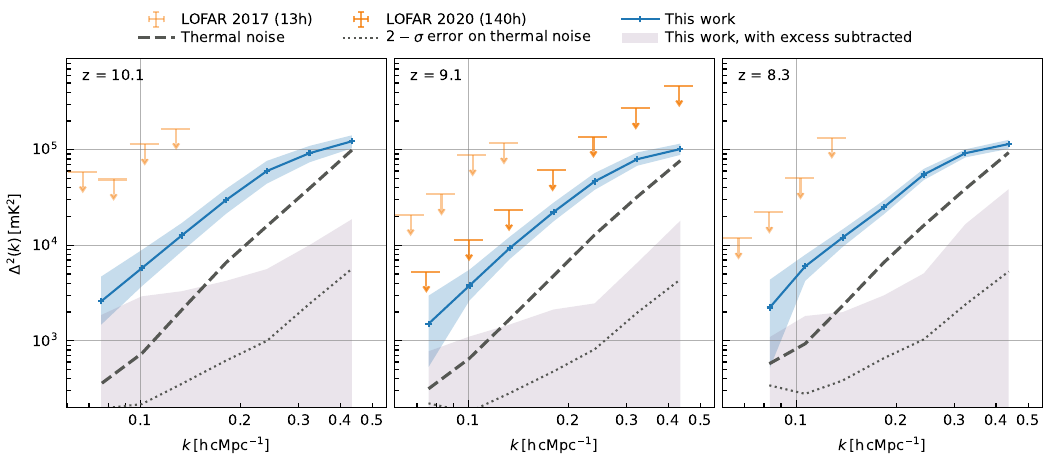}
    \caption{Final spherically averaged power spectra from the combined 10-night dataset, after ML-GPR residual foreground removal and noise bias removal, are shown as the blue line, with the blue shaded area representing the 95\% confidence interval. Previous LOFAR upper limits from \citetalias{Mertens20} (dark orange line) and \citet{Patil17} (light orange line) are included for comparison. The theoretical thermal noise power spectrum is depicted by the dashed black line, and its 2-$\sigma$ error is indicated by the dotted line. The new upper limits at $z \approx 9.1$ represent an improvement by a factor of 2 to 4 compared to the LOFAR 2020 results, achieved with comparable observational data. The lavender-grey shaded area represents the 2-$\sigma$ upper limits that would be obtained if both the `excess' and `foreground' components were subtracted from the data.}
    \label{fig:ps3d_final}
\end{figure*}

\begin{table*}
    \centering
    \caption{$\Delta_{21}^2$ upper limits at the 2-$\sigma$ level ($\Delta_{21,\mathrm{UL}}^2$) and the residual power after GPR minus the noise power ($\Delta_{I}^2 - \Delta_{N}^2$) from the 10-night dataset, at given $k$ bins for the three redshift ranges. For comparison, the~\citetalias{Mertens20} upper limits are also provided for $z \approx 9.1$.}
    \label{tab:upper_limit}
    \begin{threeparttable}
        \begin{tabular}{rrr|rrrr|rrr}
            \toprule
            \multicolumn{3}{c|}{$z\approx10.1$} & \multicolumn{4}{c|}{$z\approx9.1$} & \multicolumn{3}{c}{$z\approx8.3$} \\
            \cmidrule(lr){1-3} \cmidrule(lr){4-7} \cmidrule(lr){8-10}
            \multicolumn{1}{c}{$k$} & \multicolumn{1}{c}{$\Delta_{I}^2 - \Delta_{N}^2$} & \multicolumn{1}{c|}{$\Delta_{21,\mathrm{UL}}^2$} & \multicolumn{1}{c}{$k$} & \multicolumn{1}{c}{$\Delta_{I}^2 - \Delta_{N}^2$} & \multicolumn{1}{c}{$\Delta_{21,\mathrm{UL}}^2$} & \multicolumn{1}{c|}{$\Delta_{21,\mathrm{UL}}^2$ (2020)} & \multicolumn{1}{c}{$k$} & \multicolumn{1}{c}{$\Delta_{I}^2 - \Delta_{N}^2$} & \multicolumn{1}{c}{$\Delta_{21,\mathrm{UL}}^2$} \\
            \multicolumn{1}{c}{($h$\,cMpc$^{-1}$)} & \multicolumn{1}{c}{(mK$^2$)} & \multicolumn{1}{c|}{(mK$^2$)} & \multicolumn{1}{c}{($h$\,cMpc$^{-1}$)} & \multicolumn{1}{c}{(mK$^2$)} & \multicolumn{1}{c}{(mK$^2$)} & \multicolumn{1}{c|}{(mK$^2$)} & \multicolumn{1}{c}{($h$\,cMpc$^{-1}$)} & \multicolumn{1}{c}{(mK$^2$)} & \multicolumn{1}{c}{(mK$^2$)} \\
            \cmidrule(lr){1-10}
            0.076 & $(50.7)^2$ & $(68.7)^2$ & 0.076 & $(39.7)^2$ & $(54.3)^2$ & $(72.86)^2$ & 0.083 & $(47.4)^2$ & $(65.5)^2$ \\
            0.101 & $(75.5)^2$ & $(95.6)^2$ & 0.101 & $(62.1)^2$ & $(74.5)^2$ & $(106.65)^2$ & 0.106 & $(77.6)^2$ & $(89.5)^2$ \\
            0.133 & $(112.4)^2$ & $(130.8)^2$ & 0.133 & $(96.8)^2$ & $(110.6)^2$ & $(153.00)^2$ & 0.138 & $(109.9)^2$ & $(121.4)^2$ \\
            0.181 & $(172.4)^2$ & $(197.1)^2$ & 0.181 & $(150.1)^2$ & $(167.2)^2$ & $(246.92)^2$ & 0.184 & $(158.6)^2$ & $(171.7)^2$ \\
            0.240 & $(243.4)^2$ & $(276.2)^2$ & 0.240 & $(217.3)^2$ & $(238.6)^2$ & $(370.18)^2$ & 0.242 & $(234.5)^2$ & $(250.9)^2$ \\
            0.323 & $(302.8)^2$ & $(331.5)^2$ & 0.319 & $(283.6)^2$ & $(305.4)^2$ & $(520.33)^2$ & 0.323 & $(303.2)^2$ & $(318.2)^2$ \\
            0.434 & $(351.2)^2$ & $(376.0)^2$ & 0.432 & $(320.3)^2$ & $(341.0)^2$ & $(683.20)^2$ & 0.436 & $(338.4)^2$ & $(356.9)^2$ \\
            \bottomrule
        \end{tabular}
    \end{threeparttable}
\end{table*}

\subsection{New upper limits}

Finally, the noise bias subtracted Stokes-I spherically averaged power spectra for each of the three redshift bins ($z \approx 10.1, 9.1, \text{ and } 8.3$) is computed to derive upper limits on the 21-cm signal. The power spectra are calculated within seven $k$-bins, logarithmically spaced between $k_\mathrm{min} = 0.06\,h\,\mathrm{cMpc}^{-1}$ and $k_\mathrm{max} = 0.5\,h\,\mathrm{cMpc}^{-1}$, with a bin size of $\Delta k / k \approx 0.3$. This choice of $k$-bins balances the need for sufficient $k$-space resolution with the statistical requirements for averaging. The same input data set for the noise, derived from time-differenced visibilities and used in ML-GPR, is used to compute the power-spectra noise bias $\Delta^2_{N}$.  This is subtracted from the residual Stokes-I power, after ML-GPR, $\Delta^2_{I}$.

To estimate the uncertainties and upper limits on the 21-cm power spectrum, we recall that we employ a sampling approach inherent to our ML-GPR framework. Specifically, we generate multiple realisations of the power spectrum by sampling from the posterior distribution of the parameters obtained during the GPR analysis (see~\autoref{sec:ml_gpr}). For each realisation, we compute the spherically averaged power spectrum, incorporating the uncertainties from both the parameters and the intrinsic sampling variance. The noise power spectrum, which is subtracted from the residual power spectrum as a bias, has its own uncertainty due to sampling variance, which is also accounted for. This ensemble of power spectra allows us to construct a distribution for each $k$-bin, from which we extract the 95\% confidence interval to establish 2-$\sigma$ uncertainties and upper limits.

The resulting power spectra for the three redshift bins are presented in~\autoref{fig:ps3d_final}. Our analysis yields the most stringent 2-$\sigma$ upper limits to date on the 21-cm power spectrum from LOFAR for all three redshift bins. The new upper limits at $z \approx 9.1$ represent an improvement by a factor of 2 to 4 compared to the~\citetalias{Mertens20} results, achieved with comparable observational data. The upper limits for $z \approx 8.3$ and $z \approx 10.1$ also set new benchmarks, expanding the redshift range over which meaningful constraints can be placed on the 21-cm signal.

Despite all the improvements, the measured power spectra remain dominated by residual foregrounds and systematics, particularly at low $k$-values. These residuals are primarily due to the excess component identified in our ML-GPR decomposition. Moreover, the residuals are only partially correlated between nights, whereas a true 21-cm signal would be fully correlated (assuming it dominates over the noise), and, to a first order, isotropic (i.e., exhibiting constant power across all modes of a given $k$). Consequently, we do not interpret the positive values of $\Delta_{21}^2$ as a detection of the 21-cm signal. Instead, we conservatively treat them as upper limits.

The 2-$\sigma$ upper limits for each $k$-bin and redshift are reported in~\autoref{tab:upper_limit}. These limits are derived from the 95\% confidence intervals of the sampled power spectra distributions, accounting for both statistical uncertainties and residual systematics in the data. The most stringent 2-$\sigma$ upper limit is found for the redshift bin $z \approx 9.1$, with $\Delta_{21}^2 < (54.3\,\mathrm{mK})^2$ at $k = 0.076\,h\,\mathrm{cMpc}^{-1}$. Overall, we found that the redshift bin $z \approx 9.1$ provided the best upper limits, as it is the cleanest among the three bins analysed. This result is expected due to higher foreground levels at $z \approx 10.1$, which complicate the isolation of the 21-cm signal, and lower sensitivity at $z \approx 8.3$, caused by increased contamination from RFI at small baselines. The higher RFI contamination leads to a greater flagging fraction of the data at $z \approx 8.3$, reducing the effective sensitivity (see \autoref{fig:vis_flagger}). Nevertheless, the availability of upper limits across three distinct redshift bins enhances our ability to constrain astrophysical parameters related to the EoR, which we will present in an accompanying paper~(Ghara~et~al.,~submitted).

Following the approach of \citet{Acharya24a}, we investigated the potential improvement in the upper limits if we could cleanly isolate the `excess' component from the 21-cm signal component and subtract it from the data. The spherically averaged power spectra resulting from this hypothetical subtraction are depicted as the lavender-grey shaded area in \autoref{fig:ps3d_final}. This would yield a best 2-$\sigma$ upper limit at $z \approx 9.1$ of $\Delta_{21}^2 < (28.4,\mathrm{mK})^2$ at $k = 0.076\,h\,\mathrm{cMpc}^{-1}$. However, given our current inability to fully separate the `excess' component from the 21-cm signal component, we have decided, as in \citetalias{Mertens20}, to conservatively include the `excess' component in the final results.

Recent results from other experiments provide useful context for interpreting our new limits. The best published MWA upper limit, $\Delta_{21}^2 < (43.9~\mathrm{mK})^2$ at $z \approx 6.5$ and $k = 0.15~h\,\mathrm{cMpc^{-1}}$~\citep{Trott20}, lies at a lower redshift than the LOFAR measurements presented here. The most recent HERA upper limits~\citep{HERACollaboration23}, by contrast, overlap more directly in redshift. At $z \approx 7.9$, HERA reports a 2-$\sigma$ upper limit of $\Delta_{21}^2 < (21.4~\mathrm{mK})^2$ at $k = 0.34~h\,\mathrm{cMpc^{-1}}$, which is significantly lower than our LOFAR limit at $z \approx 8.3$, measured at a lower $k$-mode. At $z \approx 10.4$, HERA obtains a limit of $\Delta_{21}^2 < (59.1~\mathrm{mK})^2$ at $k = 0.36~h\,\mathrm{cMpc^{-1}}$, which is comparable to our result of $\Delta_{21}^2 < (68.7~\mathrm{mK})^2$ at $z \approx 10.1$, again at a lower $k$-mode.

\section{Data analysis validations}
\label{sec:validations}

Ensuring the integrity of the 21-cm signal throughout our data processing pipeline is a major concern. Given the faintness of the expected 21-cm signal and the overwhelming presence of foregrounds and instrumental effects, each step in the pipeline has the potential to inadvertently suppress or bias the 21-cm signal. In this section, we present the validation procedures applied to critical components of our pipeline. 

\subsection{Signal retention during calibration}
\label{sec:validations_calib}

The intensity scale of the visibilities, and hence ultimately the 21-cm signal strength, is fixed during the DI-calibration step. The absolute intensity scale of the sky model was set using the same procedure as described in \citetalias{Mertens20}, using the flat-spectrum source NVSSJ011732+892848 (RA01h,17m,33s, Dec~$+89^\circ,28',49''$ in J2000) with an intrinsic flux of 8.1 Jy with 5\% accuracy~\citep{Patil17}. The accuracy of our flux scale calibration was previously tested in \citetalias{Mertens20} by cross-identifying the brightest sources within $3^\circ$ of the phase centre with the 6C and 7C 151~MHz radio catalogs \citep{Baldwin85, Hales07}, confirming the reliability of our calibration. Since our DI-calibration procedure is not fundamentally changed, we refer the reader to \citetalias{Mertens20} for more detailed validation results of this step.

When subtracting the sky-model multiplied with the direction-dependent calibrated gains, there is a risk of signal suppression. \cite{Sardarabadi19} and~\cite{Mevius22} investigated the effect of DD-calibration and sky-model subtraction on signal suppression and residual noise in the power spectrum. Using the spectral smoothness constraint as implemented in \textsc{Sagecal-CO}~\citep{Yatawatta15},  overfitting and therefore signal suppression is significantly reduced. It was found that no signal suppression occurs if the baselines used in the power spectrum, i.e. those $< 250 \lambda$, are excluded from the calibration step. Excluding baselines from the calibration will result in extra noise on those baselines due to overfitting~\footnote{Overfitting occurs on the longer baselines and those errors are transferred to the baselines not used during calibration.}. However, the current smoothness constraint settings in \textsc{Sagecal-CO} significantly reduce this additional noise. \cite{Brackenhoff24} have shown that additional ionospheric noise can also be mitigated using this method. 

Signal suppression due to DI-calibration is expected to be negligible, as no visibility subtraction is involved in this step. This justifies the broader baseline selection between 50 and 5000 $\lambda$ at this stage. However, calibration errors caused by unmodelled sources and overfitting during DI-calibration can still result in additional noise~\citep{Hofer25}. This noise cannot be removed later in the calibration chain because the visibilities are multiplied by the inverse calibration gains. This effect is mitigated by minimising the number of free parameters, as we currently do by starting with a high time resolution, spectrally smooth calibration, followed by a 4-hour long time interval bandpass calibration.

We conclude that excluding short baselines from the calibration process prevents suppression of the 21-cm signal. A slight increase in variance may result from the transfer of minor gain errors from longer to shorter baselines. Nevertheless, this effect is expected to be minimal due to regularization, as shown by~\cite{Sardarabadi19,Mevius22}.

\subsection{Signal retention during RFI flagging}
\label{sec:validations_rfi}

\begin{figure}
    \includegraphics{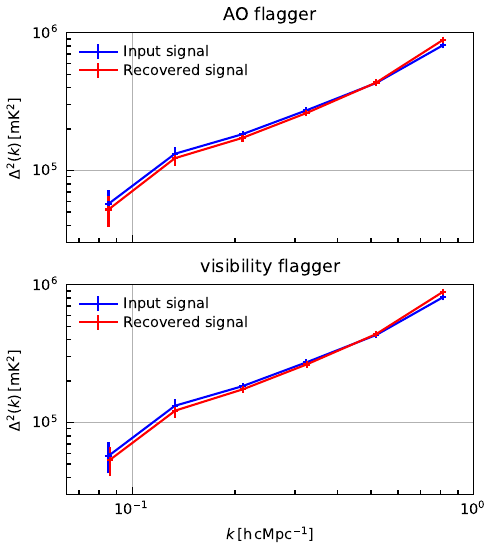}
    \caption{Recovery test of injected 21-cm signal, validating the post-processing flagging step. Top: effect after applying AOFlagger, bottom: effect of the visibility flagging step. The difference between injected and recovered signal is within the 1 sigma uncertainty. }
    \label{fig:injection_test_rfi}
\end{figure}

In order to verify that the post-calibration flagging steps  (described in \ref{sec:post_cal_flagging}) would not suppress a 21-cm signal, we inject an artificial signal in the residual visibilities of one of the observations. The power spectra after flagging with and without injected signal are subtracted to obtain a measure of the recovered signal. This is then compared to the power spectrum of the injected signal. \autoref{fig:injection_test_rfi} shows that the injected signal is recovered well, both after the wide-field AOFlagger step and after the visibility flagging. The small changes, all within the 1-$\sigma$ errors, between the injected and recovered signal are due to slight variations in the visibility standard deviation, which cause slightly different time and frequency bins to be flagged when a 21-cm signal is present. As expected, this effect decreases with decreasing amplitude of the injected signal. In order to be able to test the recovery of the injected signal over the thermal noise, we injected a signal with an amplitude larger than expected for any real 21-cm signal. We therefore conclude that the post-calibration flagging does not have a significant effect on recovery of the 21-cm signal.

\subsection{Signal retention during residual foregrounds removal}
\label{sec:validations_gpr}

\begin{figure*}
    \includegraphics[trim={0 0.15cm 0 0.15cm},clip]{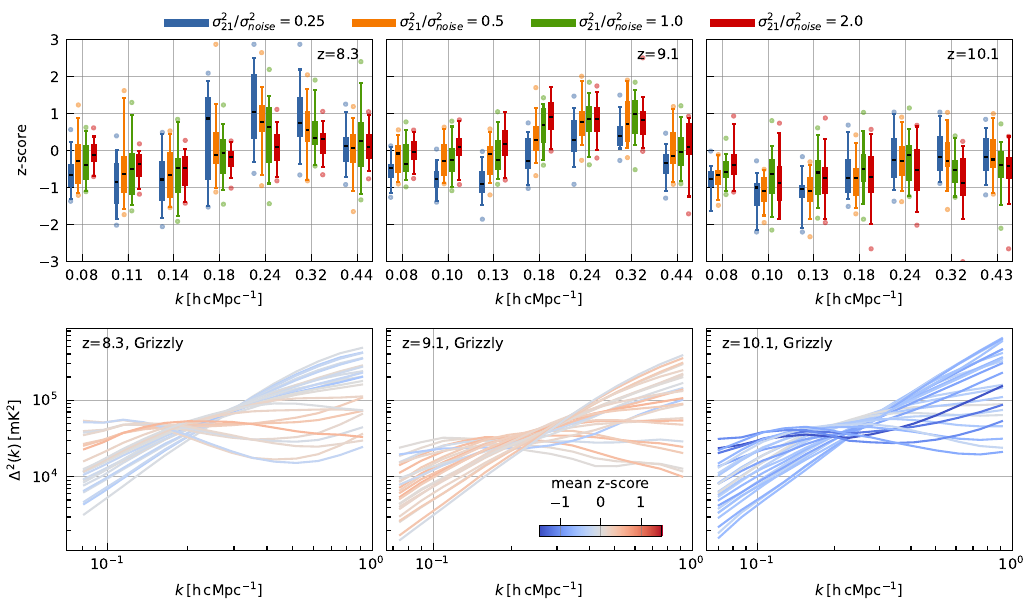}
    \caption{Results of the injection test, validating the ML-GPR residual foreground removal step, with the \textsc{Grizzly} simulations. The top plots show the z-score as a function of $k$-mode. A total of 25 synthetic 21-cm signals are tested, each with 4 different intensities. Each box-plot represents the distribution of z-scores for all 25 cases, with different box-plot colours indicating different signal intensities. The central line represents the median z-score, the box edges indicate the 25th and 75th percentiles (interquartile range), the whiskers extend to the data points within 1.5 times the interquartile range, and individual points beyond the whiskers represent outliers. A negative z-score suggests absorption of the signal, with values below $-2$ indicating absorption beyond the 2-$\sigma$ upper limits. The bottom panel shows the spherically averaged power spectrum (for $\sigma^2_{21} / \sigma^2_{\rm noise} = 1 $) of the injected signal, with colours indicating the mean z-score. }
    \label{fig:injection_grizzly}
\end{figure*}

\begin{figure*}
    \includegraphics[trim={0 0.15cm 0 0.15cm},clip]{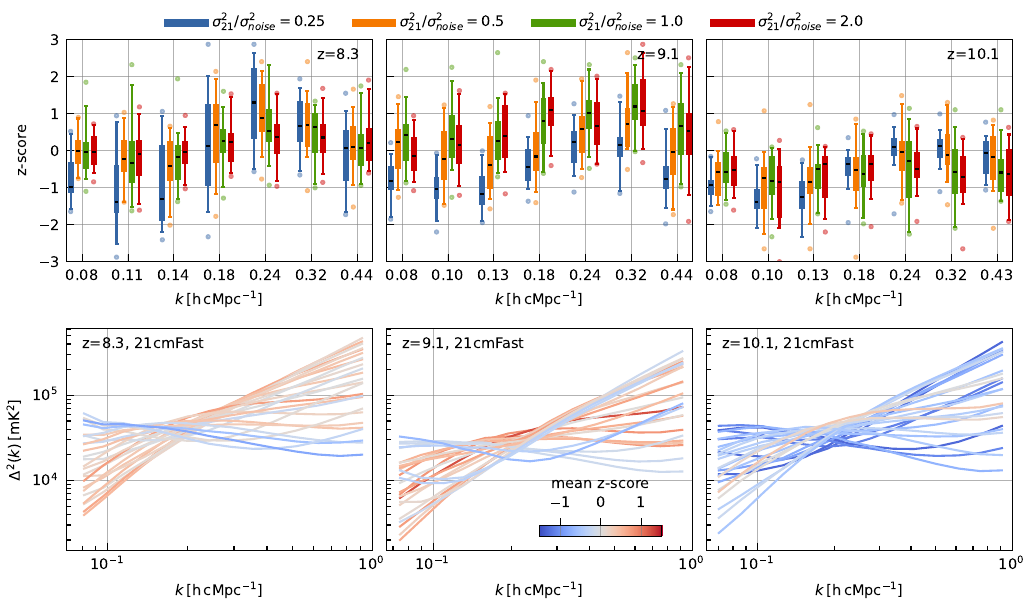}
    \caption{Same as \protect\autoref{fig:injection_grizzly} with the \textsc{21cmFAST} simulations.}
    \label{fig:injection_21cmfast}
\end{figure*}

The ML-GPR algorithm is a powerful tool for foreground mitigation, but it may inadvertently alter the recovered 21-cm signal. Assessing the impact of ML-GPR on the 21-cm signal is therefore essential to ensure the reliability of our results. To this end, we perform signal injection tests, similar to those in~\citetalias{Mertens20}, though here we employ a much more realistic and broader set of possible 21-cm signals.

Synthetic 21-cm signals are added to the data just before the ML-GPR step. The signals are generated using the decoder of the trained VAE kernel, which allows us to produce a variety of power spectra corresponding to different points in the latent space of the VAE model. We select 25 different shapes of the 21-cm power spectrum by choosing points that cover the full distribution of the training points in the latent space. This sampling captures a wide range of possible 21-cm signal shapes as represented by the \textsc{Grizzly} simulations used to train the VAE kernel. For each of these 25 shapes, we scale the power to be equal to $0.25$, $0.5$, $1.0$, and $2$ times the noise power, resulting in 100 different injected signals that cover a broad range of signal strengths. We repeated this exercise with synthetic 21-cm signals generated using a VAE kernel trained with \textsc{21cmFAST} simulations, while still using the \textsc{Grizzly} trained VAE kernel in ML-GPR. This ensures that the injection test is not dependent on the training set of the VAE kernel.

\begin{figure*}
    \includegraphics{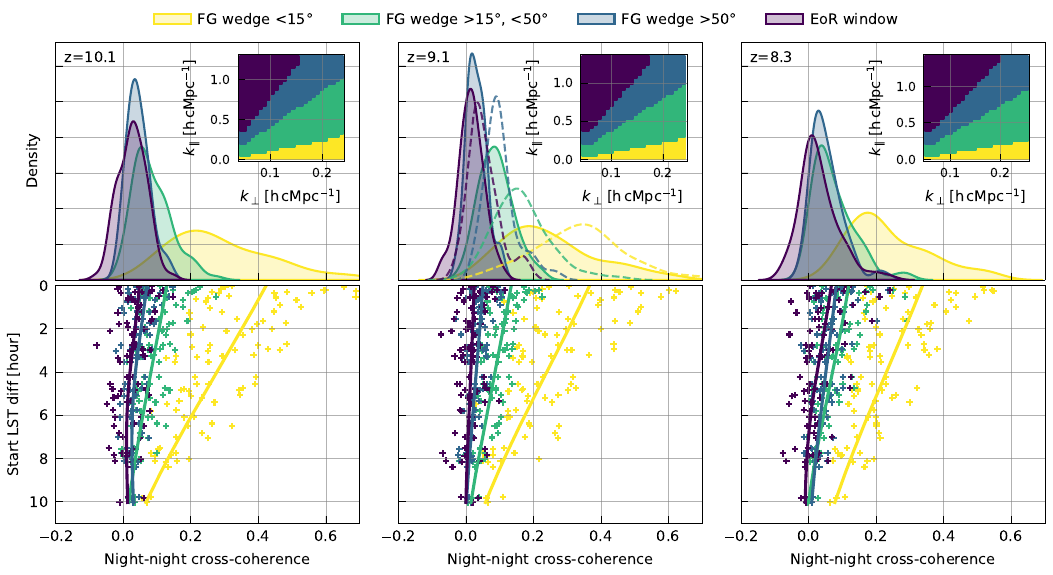}
    \caption{Night-to-night cross-coherence for different regions of the cylindrically averaged power spectra, evaluated over all pairs of nights and across the three redshift bins. The top panels display density plots of the cross-coherence values, with colours indicating different regions of the power spectra (the specific regions are illustrated in the insets). The bottom panels present the individual cross-coherence values as a function of Local Sidereal Times (LST) time difference, along with a fitted trend line to highlight the overall pattern. For the redshift bin at $z \approx 9.1$ (middle panel), we also include in dashed line the cross-coherence results from \citetalias{Mertens20} for comparison.}

    \label{fig:night_night_correlation}
\end{figure*}

For each injected signal, we generate a realisation of a visibility cube which has the desired power spectrum and add it to the data before ML-GPR. ML-GPR is then performed on this data cube with the injected signal, using the same priors as were used for the actual data without injection. The residual power spectrum of the data without an injected signal is subtracted from the residual power spectrum obtained after ML-GPR on the data with the injected signal. This difference yields the recovered 21-cm power spectrum, which can be compared to the power spectrum of the injected signal to assess any potential absorption or biasing caused by ML-GPR.

To quantify the performance of ML-GPR in recovering the injected 21-cm signal, we calculate a p-value for each $k$-bin, defined as the fraction of realisations where the residual power spectrum is higher than the injected power spectrum. We then derive the z-score using the inverse cumulative distribution function of the normal distribution. A z-score indicates the number of standard deviations by which the recovered 21-cm signal deviates from the injected 21-cm signal. A negative z-score suggests absorption of the signal, with values below $-2$ indicating absorption beyond the 2-$\sigma$ upper limits.

The z-scores for all signal injection tests are shown in the top panels of \autoref{fig:injection_grizzly} for the \textsc{Grizzly} simulation sets and \autoref{fig:injection_21cmfast} for the \textsc{21cmFAST} simulation set, with different intensities of the injected signal denoted by different colours. For the \textsc{Grizzly} simulation set, only one case (out of 100 cases) presents a few $k$-bin just below a z-score of $-2$ for the redshift bin $z\approx8.3$, none for $z\approx9.1$, and only four cases for $z\approx10.1$. For the \textsc{21cmFAST} simulation set, four cases present a few $k$-bin just below a z-score of $-2$ for the redshift bin $z\approx8.3$, one for $z\approx9.1$, and 11 for $z\approx10.1$. The lower panels of \autoref{fig:injection_grizzly} and \autoref{fig:injection_21cmfast} show the power spectra of the injected signal, with the colour indicating the mean z-score for each case. The cases with the worst recovery are those characterized by a flat power spectrum or an upturn at large scales. These cases are particularly challenging to recover as the frequency coherence scale is closer to that of the mode-mixing component. Nevertheless, ML-GPR performs remarkably well even in these challenging cases. An additional point supporting the robustness of our analysis is that, in all injection cases, the variance parameter of the 21-cm signal component ($\sigma^2_{21}$) was consistently constrained, whereas it remains unconstrained in the actual data. This indicates that if a 21-cm signal were present above the noise level, it would indeed be detected and constrained by our ML-GPR model.

Our signal injection tests demonstrate that ML-GPR does not significantly suppress the 21-cm signal. The recovered power spectra match the injected signals within the uncertainties for almost all cases, confirming the reliability of ML-GPR in our analysis. Over the 600 injections that we performed (3 redshift bins, 2 simulation codes), only 3.5\% presented a z-score below $-2$. We also note that the z-score for the smallest $k$-bins was consistently above $-2$, indicating robust recovery for the larger spatial scales. When considering individual $k$-bins, only 1.1\% of the bins exhibited a z-score below $-2$. This validation shows that our ML-GPR pipeline preserves the 21-cm signal while effectively mitigating foregrounds and systematics. Furthermore, these results indicate that any potential bias in the 21-cm power spectrum is minimal and does not necessitate correction, thereby supporting the robustness of our analysis.

\section{Discussion}
\label{sec:discussion}
  
Throughout the reprocessing of the LOFAR observations originally analysed in~\citetalias{Mertens20}, we have significantly enhanced our data processing pipeline. Each step was carefully reviewed and refined to minimise sources of excess power that hinder the detection of the 21-cm signal. In this section, we review the changes made to the pipeline and how they contribute to the overall improvement, examine the sources of remaining excess power, and discuss further developments needed to enhance the quality of our results.

\subsection{Quantifying pipeline improvements}

Several key updates to the pipeline contributed to the reduction of excess variance across the dataset. The revised DI-calibration scheme, which separates spectral and temporal components, led to a modest reduction in excess variance of about 10--20 per cent. More importantly, it reduced gain error by 2 to 3 orders of magnitude, making the calibration significantly more robust. The updated DD-calibration, particularly the removal of some source clusters beyond the first beam null, reduced foreground contamination within the wedge by up to a factor of five for some nights. However, the level of improvement varied with both night and observed LST range, leading to a more modest average reduction—closer to a factor of two or less on the combined dataset. Enhanced RFI flagging after calibration reduced power by up to a factor of 10 on short baselines in some cases, though night-to-night variability means the average improvement is closer to a factor of 4 to 5. Finally, the refinement of the foreground removal algorithm using ML-GPR played a key role in further reducing the final excess variance. Its effectiveness was supported by the improved calibration and RFI flagging, which helped confine foreground power to the low $k_\parallel$ range. This improved the separation between foregrounds and residual (excess and 21-cm) components, making the foreground removal process more efficient and reliable.

\subsection{Sources of excess}

We investigated many sources of excess power during this reprocessing:

\begin{description}
  \setlength\itemsep{1em}
  
    \item[\textit{Bright sources in the side-lobes ---}] Bright sources, far outside the primary beam, such as Cas\,A and Cygnus\,A present significant challenges. \cite{Gan22} identified them as a source of excess variance in the~\citetalias{Mertens20} results. During observations, the LOFAR station beam gain in their direction evolves considerably. The deep nulls and strong spectral variations in the sidelobes of the beam, resulting from the regular structure of the array, paired with beam modelling errors, make subtracting these sources during DD-calibration very difficult, especially since we impose spectrally smooth solutions. Any residual flux from these sources introduces spectral fluctuations into the data. Additionally, ~\citet{Munshi24} showed that bright distant sources can contribute power at much higher $k_\parallel$ than previously thought. Currently, only Cas\,A and Cygnus\,A are addressed, while many other sources, such as Taurus\,A, Virgo\,A, and other bright 3C sources, may impact our data as well. This is clearly a source of excess power. Although improvements in the Cas\,A and Cygnus\,A sky models and directional post-calibration flagging helped reduce their impact, much work remains to be done to further mitigate their impact \citep{Ceccotti25}. Optimising calibration and flagging (or avoidance) strategies for bright sources in the sidelobes is one of the major topics currently under study.

    \item[\textit{DD-calibration errors ---}] Enforcing spectrally smooth solutions for DD-calibration has significantly reduced excess variance due to calibration errors. However, related to the point above, we noticed that calibrating and subtracting clusters of sources located far beyond the first null of the beam may in some circumstances do more harm than good. In current analysis, removing those clusters from the sky model improved the final 21-cm signal power spectrum limit. Deciding which sources to include in the final sky model, however, requires further investigation and automation.

    \item[\textit{DI-calibration errors ---}] Improving DI-calibration was a major focus of this reprocessing, and we managed to considerably reduce the gain errors in this step. The number of free parameters in DI-calibration was significantly reduced by allowing only a time-stable but spectrally varying bandpass gain, and conversely applying time-varying but spectrally smooth solutions to the visibilities. This approach minimises the impact of DI-calibration errors on the power spectra, particularly at higher delays. We believe DI-calibration errors to not be a significant source of excess power anymore. This will be further quantified by \citep{Hofer25}. 

    \item[\textit{Radio frequency interference ---}] In the reprocessing, we observed the significant impact that low-level RFI can have on the power spectra (see e.g.~\autoref{fig:vis_flagger}); it was clearly a source of excess power. Broadband RFI can introduce frequency structure at high $k_\parallel$ and is usually difficult to detect and flag. Processing observations targeting 3C\,196, for example, showed that, for targets other than the NCP where RFI does not coherently add up, the excess is reduced (Ceccotti~et~al.,~in~prep.). While major improvements were implemented in the pipeline to reduce the impact of RFI, which have clearly been effective, further work is needed: our current approach, which discards any data -- even entire baselines -- affected by RFI, has the negative effect of decreasing our sensitivity. This is especially the case for small baselines, which are heavily affected by local sources of RFI, thereby reducing our sensitivity at the largest scales.

    \item[\textit{Ionosphere errors ---}] In a recent paper by~\cite{Brackenhoff24} it is shown that calibration errors induced by ionospheric disturbances are unlikely to be the cause of excess variance. \cite{Gan22} also did not observe any correlation between metrics assessing ionospheric activity and excess variance in the observations. However, the interaction of ionospheric errors with other effects, such as beam errors in the far sidelobes, requires further investigation.

    % \item[\textbf{Polarization}] There are no new findings regarding polarization effects, and they are most certainly not a significant concern in our current analysis.

\end{description}

\subsection{Future improvement}

\autoref{fig:night_night_correlation} effectively illustrates the improvements achieved in our processing pipeline to reduce the aforementioned sources of excess power, as well as highlighting areas that still require attention. The figure shows the night-to-night cross-coherence for different regions of the cylindrically averaged power spectra, evaluated over all pairs of nights and across the three redshift bins. We split the cylindrically averaged power-spectra into four regions, corresponding different angular ranges: (i) the region covered by our NCP sky model ($< 15^\circ$, in yellow), (ii) the region where we expect most of the power from the bright sources Cas\,A and Cygnus\,A ($15^\circ$ to $50^\circ$, in green), (iii) the foreground wedge region affected by more distant bright sources ($> 50^\circ$, light blue), and (iv) the EoR window (dark blue). Compared to the results from \citetalias{Mertens20} (indicated by the dashed line in the top-middle panel), we observe a considerable reduction in night-to-night correlation across all $k$-space regions. Specifically, the correlation in the EoR window and in the foreground wedge above $50^\circ$ is now very close to zero, indicating minimal remaining coherent excess power in these regions. However, some correlation remains in the foreground wedge within the $15^\circ$ to $50^\circ$ range, suggesting that distant and bright sources continue to contribute a coherent excess. Additionally, significant correlation persists in the foreground wedge below $15^\circ$, pointing to contributions from sources within the first few sidelobes of the LOFAR station beam. The fact that the night-to-night correlation decreases considerably when the two nights are observed at very different local sidereal times -- and thus see the sky through a very different primary beam -- as shown in the lower panel of~\autoref{fig:night_night_correlation}, indicates that these residual excesses have a sky origin. 

Most of the causes of excess power, we currently believe, could be mitigated by further improving the calibration scheme, our sky model, and the GPR covariance model. Enhancing our RFI mitigation strategy could also help in preserving our sensibility at large scale.

\begin{description}
  \setlength\itemsep{1em}

  \item[\textit{Further improving the low level RFI flagging ---}] A possibility under investigation is to use a more specific filtering instead of flagging full baselines. As some of the strong sources of RFI identified in this work are often located near the superterp, a procedure similar to a DD-calibration, but for a source on the ground, could be envisaged~\citep{Finlay23}. A similar method is also investigated for the NenuFAR data~\citep{Munshi25b}.

  \item[\textit{Improving the GPR covariance model ---}] The separation between residual foregrounds and the 21-cm signal remains a significant challenge in our analysis. Further enhancements to our ML-GPR framework are necessary to address this issue. Our strategy has been to refine the covariance model to better match the actual covariance of the data. This approach was successfully applied to the 21-cm signal component by implementing learned kernels trained on 21-cm simulations. A similar methodology needs to be adopted for the foreground covariance model, which currently relies on a generic covariance function. Developing an analytically defined or simulation-based learned foreground covariance model that incorporates instrumental and systematic effects should greatly enhance the accuracy of our foreground modelling. Additionally, extending the ML-GPR framework to enable a joint analysis over multiple redshifts simultaneously -- thereby encompassing a wider bandwidth -- could improve our ability to distinguish between the evolving 21-cm signal and the spectrally smooth foregrounds. Implementing this strategy would require our calibration scheme to be executed consistently over the combined redshift bins.

  \item[\textit{Beam model and calibration---}] For computational reasons, the beam model is until now not used in DD-calibration. Yet, we see that a major source of excess noise is related to beam effects outside the first null of the beam. Mitigating beam effects in DD-calibration will be a major focus of future improvements to our calibration scheme. This can be accomplished by smart weighting schemes that down weight the effects of nulls in the beam~\citep{Brackenhoff25} or reducing the number of degrees of freedom by spatial constraints as discussed in \cite{Yatawatta22}. Other ways of reducing the number of degrees of freedom in DD-calibration involve decomposing the calibration in steps that are physically motivated, e.g. ionospheric effects.   

 \item[\textit{Improving the NCP sky model ---}] In the current analysis, we discarded some of the outer source clusters that were previously used in the NCP sky model, because keeping them in the model would only add to the excess noise, at least for the observations we checked. This evaluation needs to be performed more rigorously. The number of clusters and components in the sky model should be revisited and possibly reduced. At the same time, we notice that bright sources that may  have a low apparent brightness when integrated over the full 12 hr observations can cause issues at certain sidereal times when they are in the sidelobes of some of the beams. Therefore, we should investigate the effect of adding more bright sources other than Cas\,A  and Cygnus\,A.

\end{description}

\section{Summary and Conclusion}

We have presented new upper limits on the 21-cm signal from the Epoch of Reionization, derived from reprocessed LOFAR observations. Building upon the work of \citetalias{Mertens20}, we have significantly enhanced our data processing pipeline and extended our analysis to a broader frequency range, allowing us to set upper limits at redshifts $z \approx 10.1, 9.1, \text{ and } 8.3$. The main conclusions of our work are:

\begin{enumerate}[leftmargin=1.5em,label=(\roman*)]
    \setlength\itemsep{1em}
    
    \item By significantly modifying our DI-calibration strategy, splitting the spectral and temporal calibration parts, a reduction in gain errors by 2 to 3 orders of magnitude was achieved by significantly reducing the number of free parameters. This minimised the impact of DI-calibration errors on the power spectra, particularly at higher delays, effectively eliminating them as a source of excess power.

    \item We updated our sky model, incorporating improved models of distant bright sources like Cas\,A and Cygnus\,A, which are known to introduce excess variance due to their complex beam interactions. Despite these improvements, residual contributions from these and other bright sources outside the primary beam remain significant sources of excess power. In addition, we found that calibrating and subtracting clusters of sources located far beyond the first null of the beam sometimes increased errors. By removing these distant clusters from the sky model, we improved the final 21-cm signal power spectrum limit.
   
    \item By implementing a new post-calibration RFI mitigation strategy, including delay-space baseline flagging, we significantly reduce excess power caused by low-level and broadband RFI, particularly from local sources. This approach effectively mitigated RFI contamination on small baselines, leading to reductions in power by factors of 4 to 5 at specific $k_{\perp}$ values. While this leads to increased thermal noise on large scale due to data loss from flagging entire baselines, the overall benefit in reducing RFI-induced excess power far outweighed the sensitivity loss, enhancing the quality of our power spectrum measurements.
        
    \item The Gaussian Process Regression (GPR) method was enhanced by employing a machine learning approach to construct a physically motivated covariance function for the 21-cm signal, thereby improving the separation between the 21-cm signal and foreground components. Additionally, we refined the foreground covariance model by making the coherence scale dependent on baseline length, effectively accounting for the foreground `wedge' in $k$-space. These advancements substantially improved our ability to isolate the 21-cm signal from foreground contamination.
        
    \item The cumulative effect of all these improvements is a significant reduction of residual power across the entire $k$-range, effectively minimising systematics in our data. At the redshift bin $z \approx 9.1$, residual power inside the wedge decreased by a factor of about two compared to our previous analysis. Specifically, the ratio of residual power to thermal noise decreased from 6.2 to 3 inside the wedge, and from approximately 3 to 1.6 outside it. These advancements bring our measurements closer to the theoretical thermal noise limit, thereby improving the sensitivity of our observations.

    \item We have established the most stringent 2-$\sigma$ upper limits on the 21-cm signal from LOFAR at redshifts $z \approx 10.1, 9.1, \text{ and } 8.3$. Specifically, at $z \approx 9.1$, we achieved a 2 to 4-fold improvement over our previous results \citepalias{Mertens20} using comparable observational data, setting a best upper limit of $\Delta_{21}^2 < (54.3\,\mathrm{mK})^2$ at $k = 0.076\,h\,\mathrm{cMpc}^{-1}$. For the other redshifts, we achieved a best upper limit of $\Delta_{21}^2 < (68.7\,\mathrm{mK})^2$ at $k = 0.076\,h\,\mathrm{cMpc}^{-1}$ for $z \approx 10.1$ and $\Delta_{21}^2 < (65.5\,\mathrm{mK})^2$ at $k = 0.083\,h\,\mathrm{cMpc}^{-1}$ for $z \approx 8.3$. The upper limits for each $k$-bin and redshift are reported in~\autoref{tab:upper_limit}. These upper limits have been rigorously validated through comprehensive tests, including signal injection, ensuring that our data processing and analysis methods do not suppress the 21-cm signal.

\end{enumerate}

These new multi-redshift upper limits provide new constraints that can be used to refine our understanding of the astrophysical processes during the EoR. The implications of these multi-redshift 21-cm signal power spectrum upper limits are presented in an accompanying paper by Ghara~et~al.~(submitted). The study uses a Bayesian inference framework based on the 21-cm signal power spectrum modelling using {\sc Grizzly} code to constrain the IGM properties of the disfavoured reionization scenarios between redshift 8--10. The study shows that the disfavoured models are still extreme types in which the 21-cm signal fluctuations are mainly driven by rare and large ionized/emission regions. For a standard cosmology scenario without any excess radio background to the CMB, the 95 per cent credible intervals of the disfavoured models at redshift 9.1 represent disfavoured IGM states with averaged ionization and heated fraction below $\lesssim 0.55$, an average gas temperature  $\lesssim 21$ K and characteristic size of the heated region $\lesssim 40$ Mpc. These constraints are based on uniform priors of the {\sc Grizzly} source parameters on their ranges and by using conservative limits on the maximum ionization fraction at those three redshifts, estimated from the CMB Thomson scattering optical depth from Planck.  We refer to Ghara~et~al.~(submitted) for the detailed constraints on the source and IGM parameters for all three redshifts.

% Their implications are analysed in detail in an accompanying paper by Ghara~et~al.~(submitted), using the reionization simulation code \textsc{Grizzly}~\citep{Ghara15,Ghara18,Ghara20} and a Bayesian inference framework to constrain the parameters of the IGM.

While we have not yet detected the 21-cm signal, the significant improvements achieved in this work lay the groundwork for future analyses with much longer ($\sim$1000 hours) integrations and more advanced data processing techniques. Continued development in RFI mitigation, calibration strategies, beam modelling, and sky modelling will further enhance our ability to detect the elusive 21-cm signal from the Epoch of Reionization.

\section*{Acknowledgements}

We are grateful to the referee for their valuable feedback and suggestions that improved this paper. FGM acknowledges the financial support of the PSL Fellowship Programme. LVEK, SAB, KC, SG, CH and SM acknowledge the financial support from the European Research Council (ERC) under the European Union’s Horizon 2020 research and innovation programme (Grant agreement No. 884760, `CoDEX'). EC (INAF) would like to acknowledge support from the Centre for Data Science and Systems Complexity (DSSC), Faculty of Science and Engineering at the University of Groningen, and from the Ministry of Universities and Research (MUR) through the PRIN project `Optimal inference from radio images of the epoch of reionization'. GM is supported by Swedish Research Council grant 2020-04691. QM acknowledges the financial support of the National Natural Science Foundation of China (Grant No. 12263002). RG acknowledges support from SERB, DST Ramanujan Fellowship no. RJF/2022/000141. EC (Nottingham) acknowledges the support of a Royal Society Dorothy Hodgkin Fellowship and a Royal Society Enhancement Award. SKG is supported by NWO grant number OCENW.M.22.307.

\section*{Data Availability}
The data underlying this article will be shared on reasonable request to the corresponding author.

%%%%%%%%%%%%%%%%%%%%%%%%%%%%%%%%%%%%%%%%%%%%%%%%%%

%%%%%%%%%%%%%%%%%%%% REFERENCES %%%%%%%%%%%%%%%%%%

% The best way to enter references is to use BibTeX:

\bibliographystyle{mnras}
\bibliography{bibliography.bib} % if your bibtex file is called example.bib

% Alternatively you could enter them by hand, like this:
% This method is tedious and prone to error if you have lots of references
%\begin{thebibliography}{99}
%\bibitem[\protect\citeauthoryear{Author}{2012}]{Author2012}
%Author A.~N., 2013, Journal of Improbable Astronomy, 1, 1
%\bibitem[\protect\citeauthoryear{Others}{2013}]{Others2013}
%Others S., 2012, Journal of Interesting Stuff, 17, 198
%\end{thebibliography}

%%%%%%%%%%%%%%%%%%%%%%%%%%%%%%%%%%%%%%%%%%%%%%%%%%

%%%%%%%%%%%%%%%%% APPENDICES %%%%%%%%%%%%%%%%%%%%%

\newpage
\appendix

\section{Example of RFI affected baselines}
\label{sec:ap_rfi_baselines}

This appendix presents a few example of baseline affected by RFI in the 147--149 MHz and 155--158 MHz range, and also a typical baseline affected by broad-band RFI and flagged by the delay-space baseline flagger.

\begin{figure}[!ht]
    \includegraphics{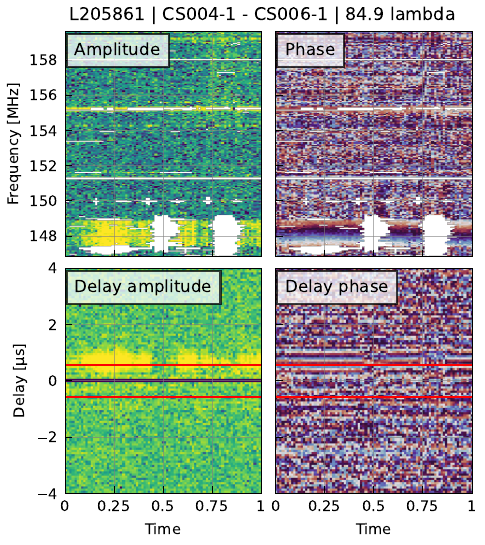}
    \caption{An example of a baseline affected by RFI in the 147--149 MHz range. The red solid line represent the baseline horizon line.}
    \label{fig:vis_flagger_ex_148}
\end{figure}

\begin{figure}[!ht]
    \includegraphics{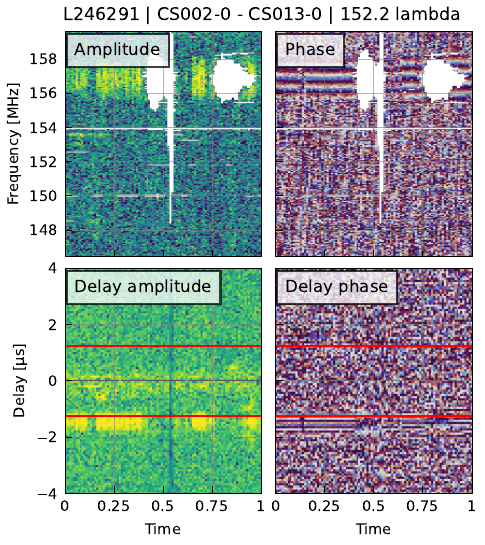}
    \caption{An example of a baseline affected by RFI in the 155--158 MHz range. The red solid line represent the baseline horizon line.}
    \label{fig:vis_flagger_ex_156}
\end{figure}

\begin{figure}[!ht]
    \includegraphics{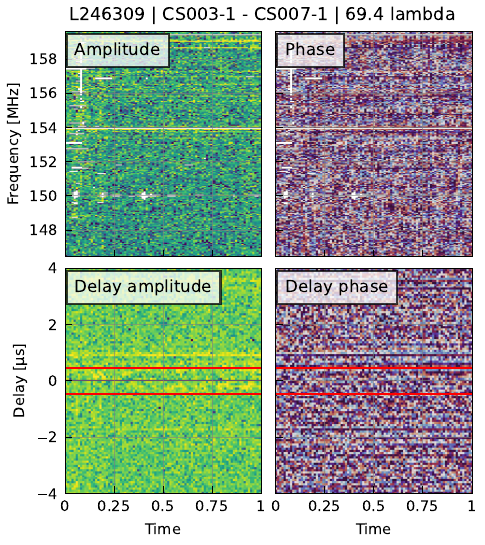}
    \caption{An example of a baseline broad-band RFI flagged by the delay-space baseline flagger. The red solid line represent the baseline horizon line.}
    \label{fig:vis_flagger_ex_delay}
\end{figure}

\onecolumn
\newpage
\section{Cylindrically-averaged power-spectra all nights}
\label{sec:ap_all_ps2d}

This appendix presents all cylindrically-averaged power-spectra of the 14 processed nights before and after residual foregrounds removal (ML-GPR).

\begin{figure*}[!ht]
    \includegraphics{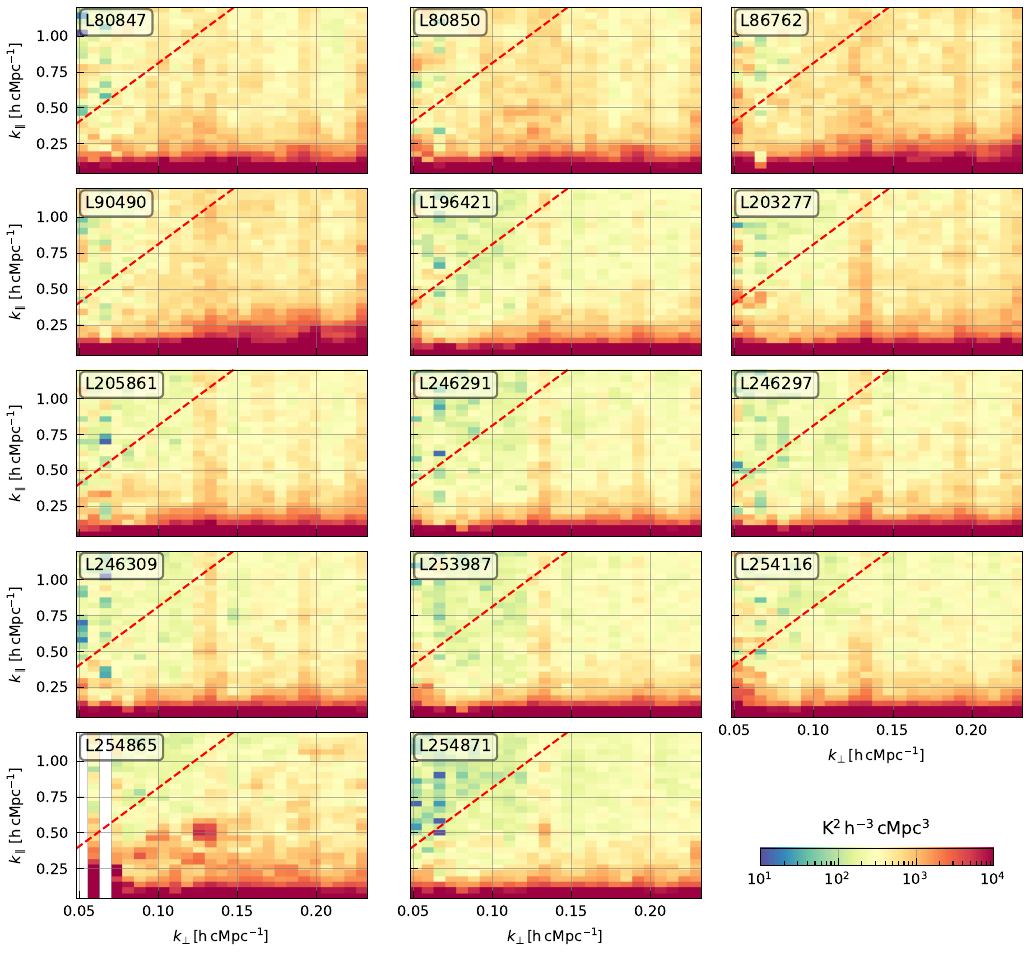}
    \caption{Cylindrically-averaged power-spectra of all nights at $z \approx 10.1$, before ML-GPR.}
    \label{fig:ps2d_i_all_nights_eor1}
\end{figure*}

\begin{figure*}[!ht]
    \includegraphics{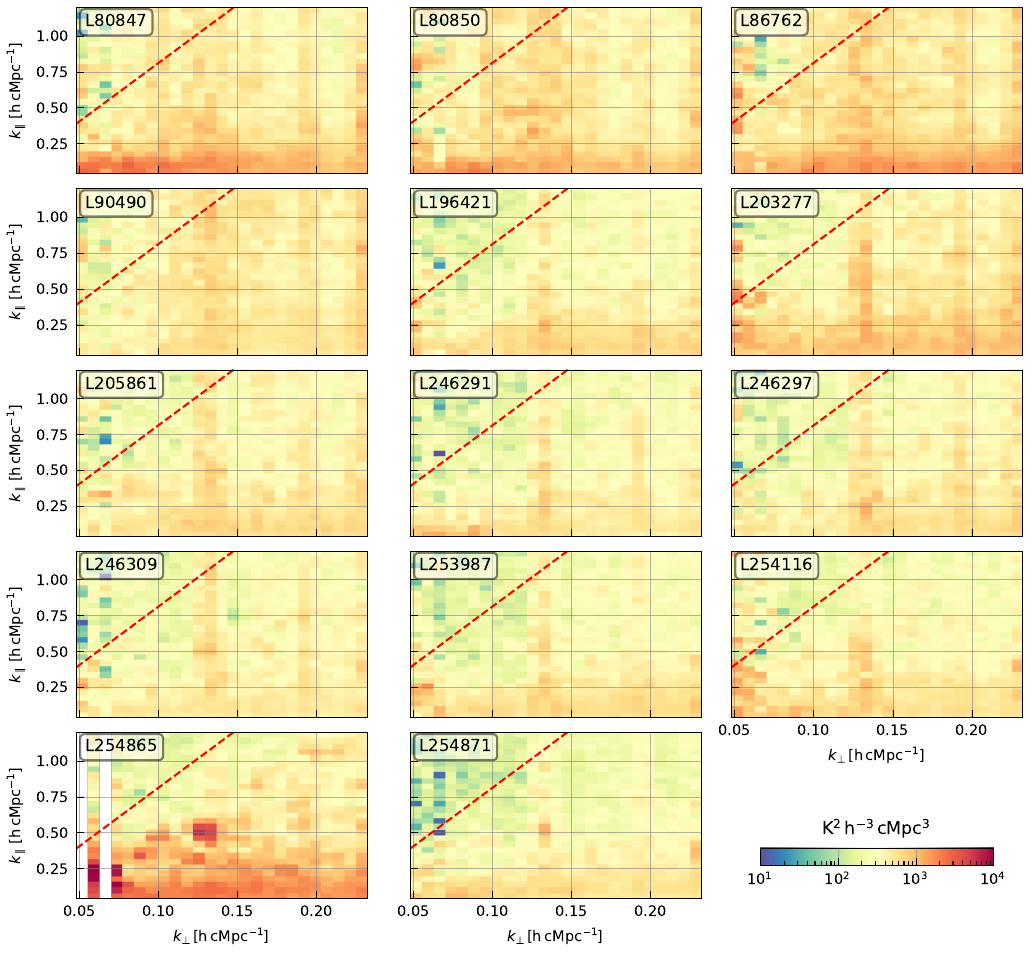}
    \caption{Cylindrically-averaged power-spectra of all nights at $z \approx 10.1$, after ML-GPR.}
    \label{fig:ps2d_res_all_nights_eor1}
\end{figure*}

\begin{figure*}[!ht]
    \includegraphics{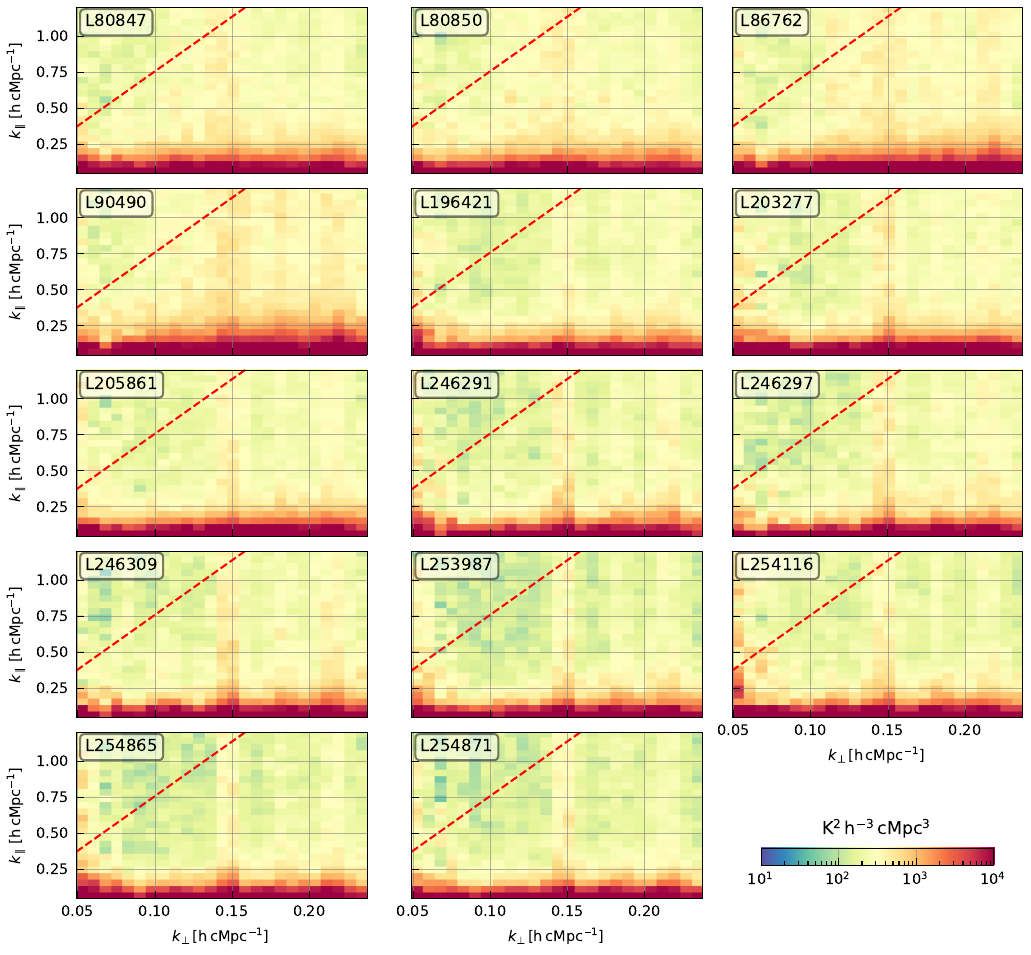}
    \caption{Cylindrically-averaged power-spectra of all nights at $z \approx 9.1$, before ML-GPR.}
    \label{fig:ps2d_i_all_nights_eor2}
\end{figure*}

\begin{figure*}[!ht]
    \includegraphics{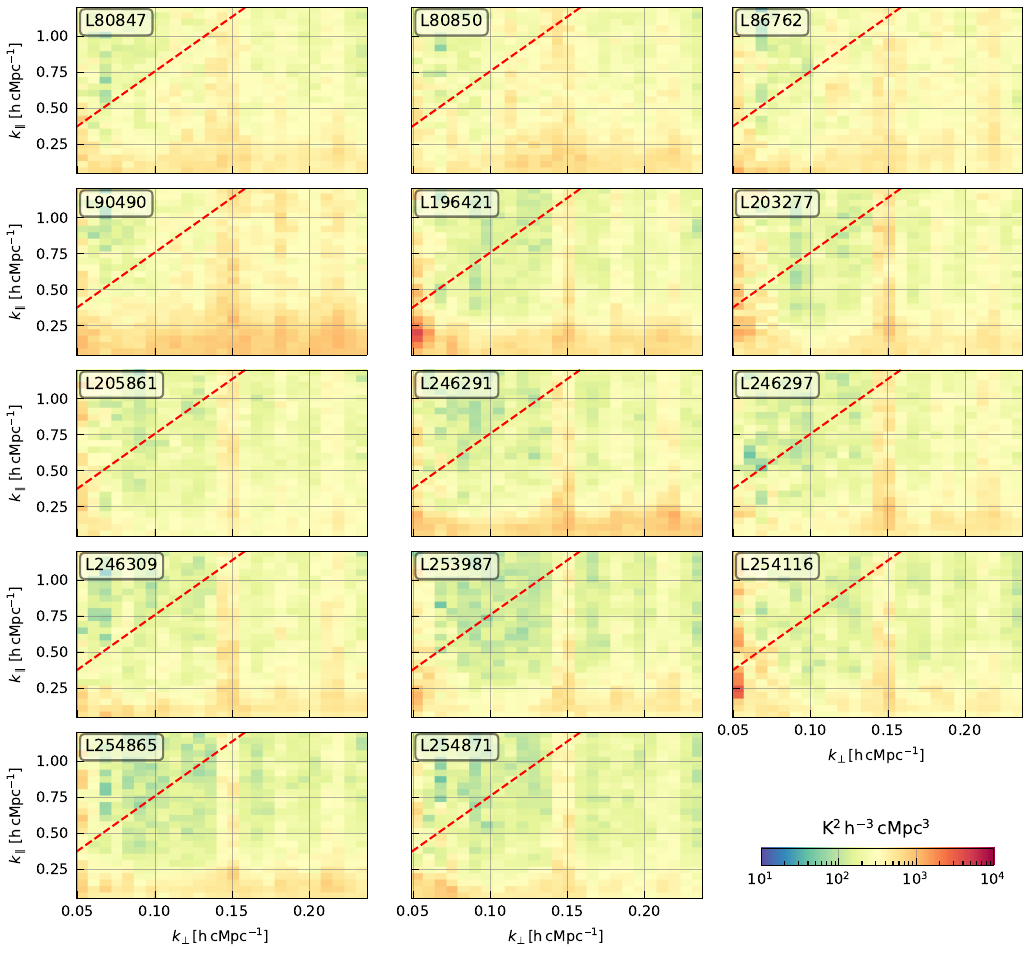}
    \caption{Cylindrically-averaged power-spectra of all nights at $z \approx 9.1$, after ML-GPR.}
    \label{fig:ps2d_res_all_nights_eor2}
\end{figure*}

\begin{figure*}[!ht]
    \includegraphics{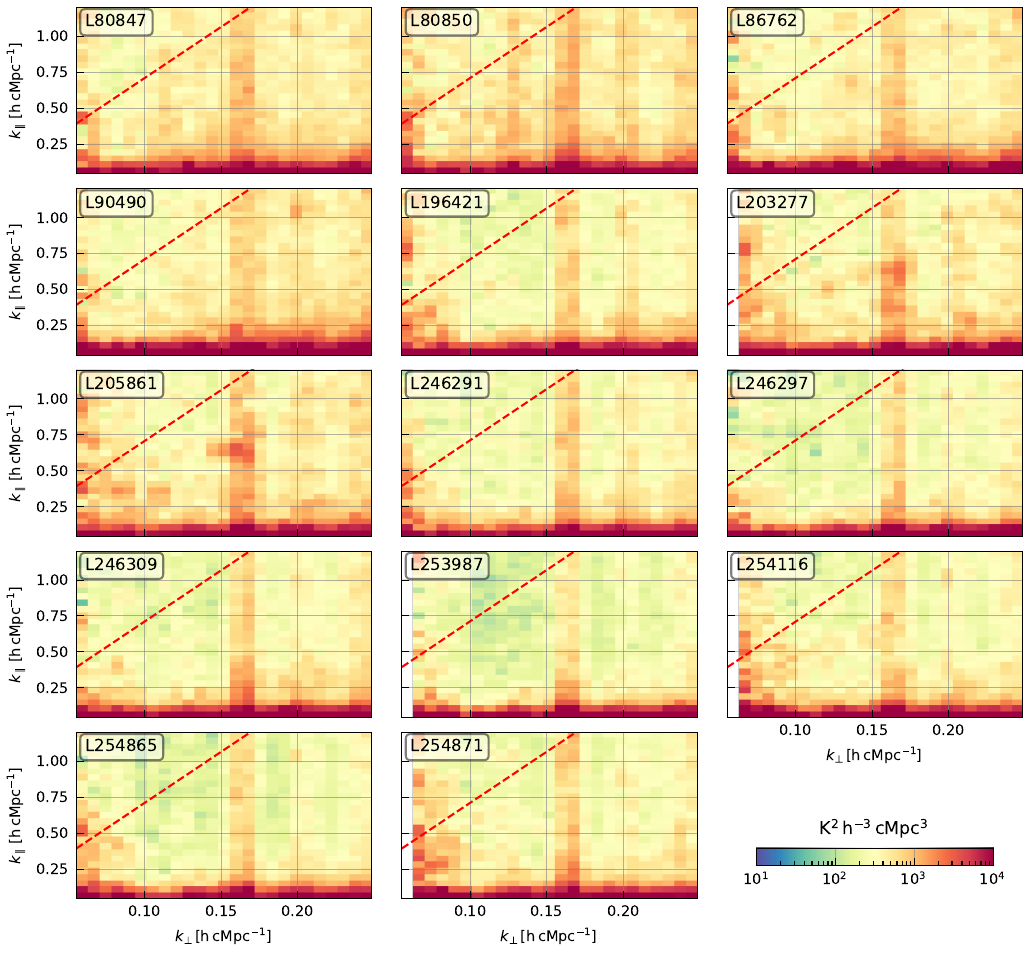}
    \caption{Cylindrically-averaged power-spectra of all nights at $z \approx 8.3$, before ML-GPR.}
    \label{fig:ps2d_i_all_nights_eor3}
\end{figure*}

\begin{figure*}[!ht]
    \includegraphics{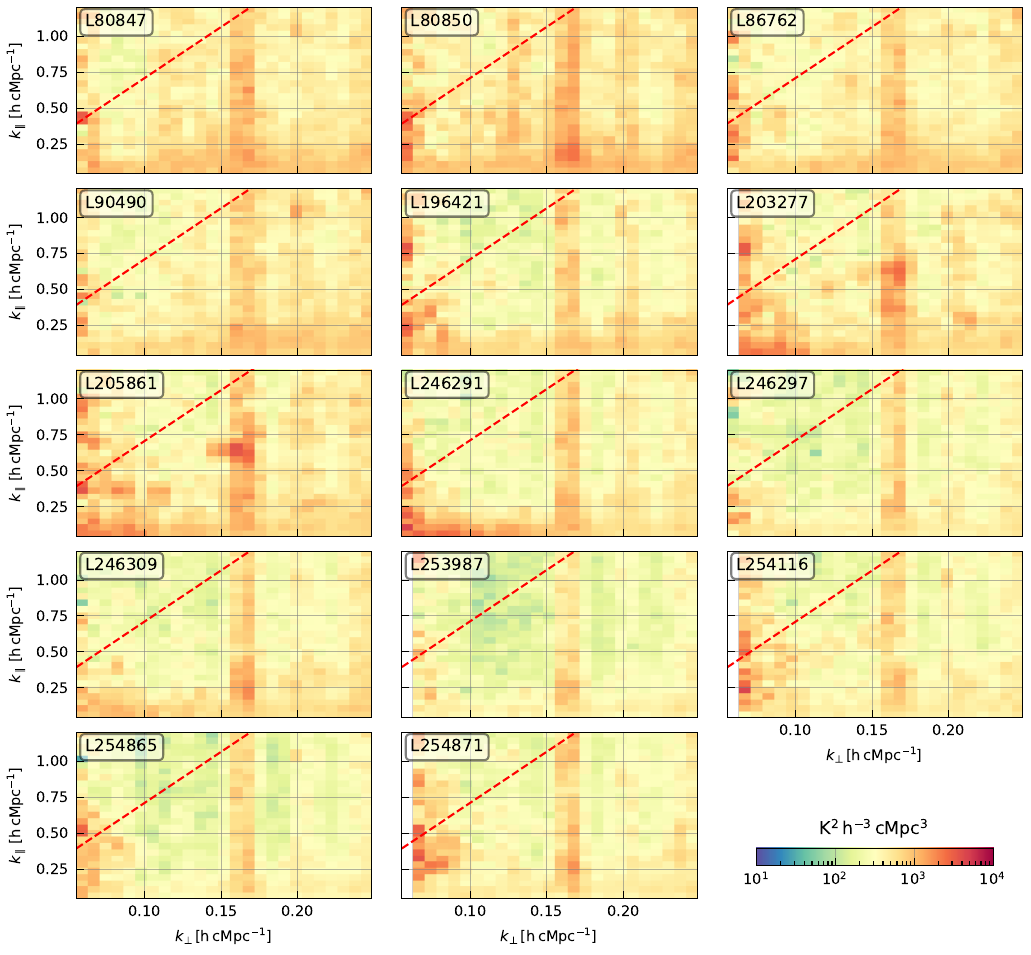}
    \caption{Cylindrically-averaged power-spectra of all nights at $z \approx 8.3$, after ML-GPR.}
    \label{fig:ps2d_res_all_nights_eor3}
\end{figure*}

\newpage

\section{Posterior distribution of the GP model hyper-parameters}
\label{sec:ap_all_corner}

This appendix presents all posterior distribution of the Gaussian Process model parameters for all three redshift bins, using a nested sampling algorithm.

\begin{figure*}[!ht]
    \includegraphics{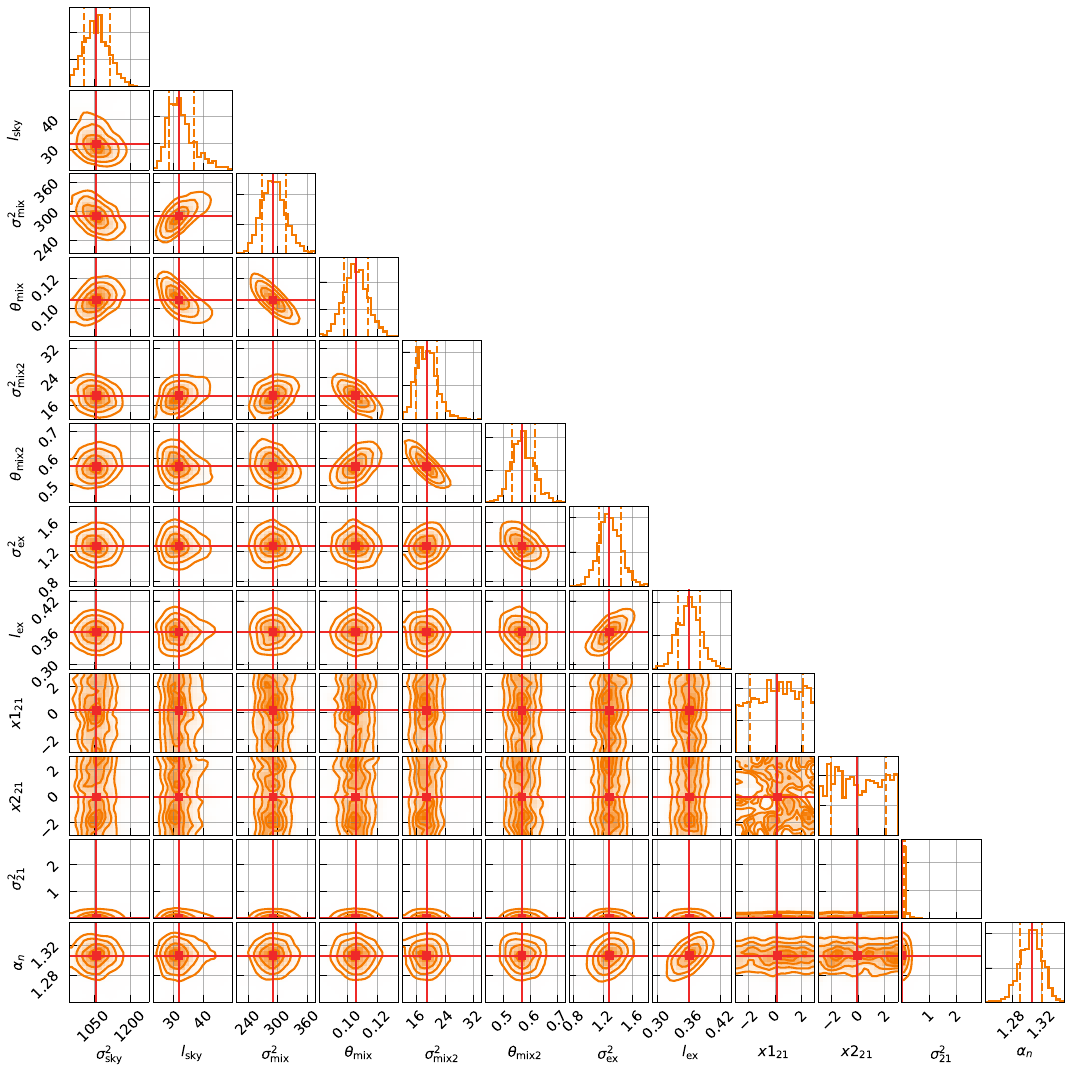}
    \caption{Posterior distribution of the Gaussian Process model parameters derived using a nested sampling algorithm, at $z \approx 10.1$}
    \label{fig:gpr_corner_eor1}
\end{figure*}

\begin{figure*}[!ht]
    \includegraphics{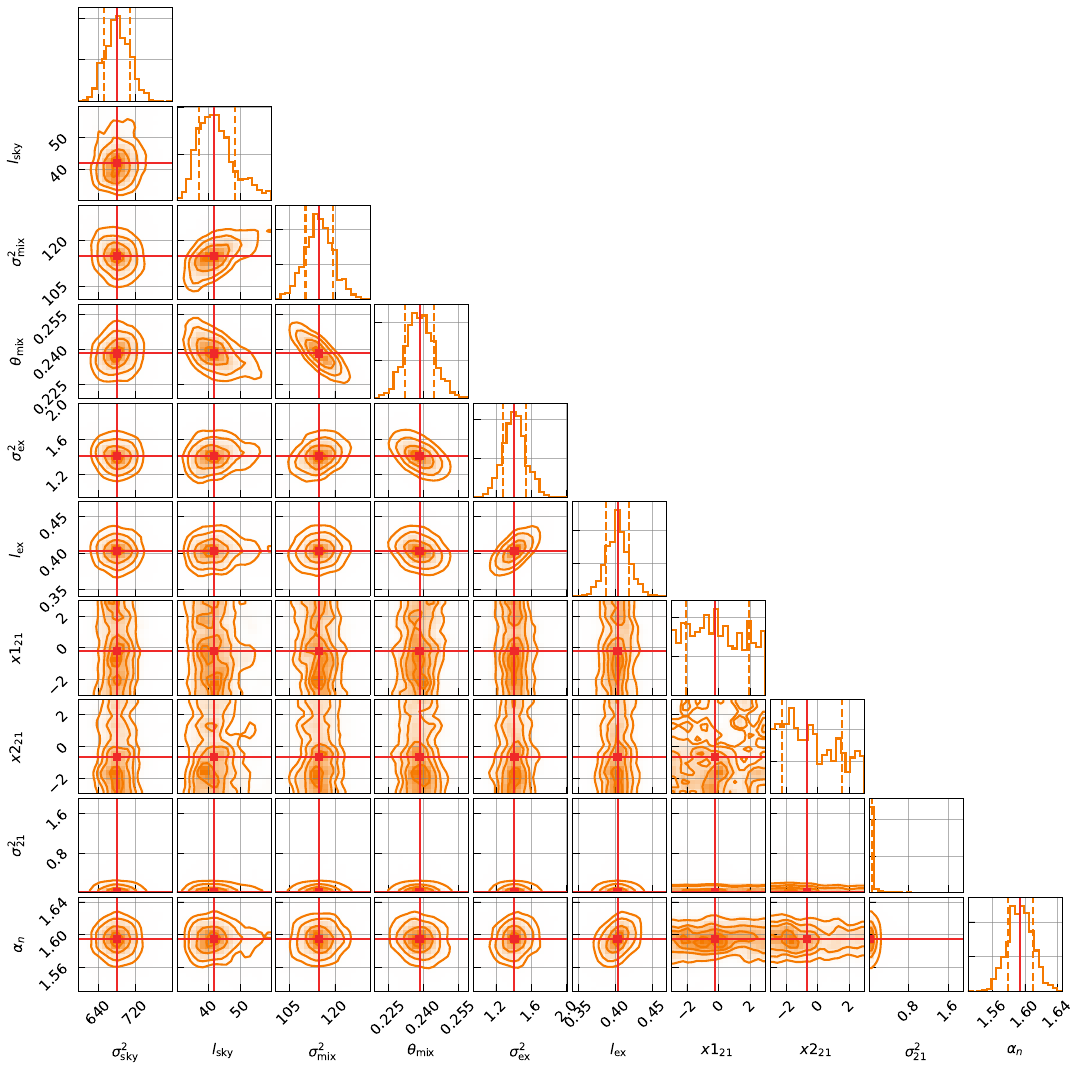}
    \caption{Posterior distribution of the Gaussian Process model parameters derived using a nested sampling algorithm, at $z \approx 9.1$}
    \label{fig:gpr_corner_eor2}
\end{figure*}

\begin{figure*}[!ht]
    \includegraphics{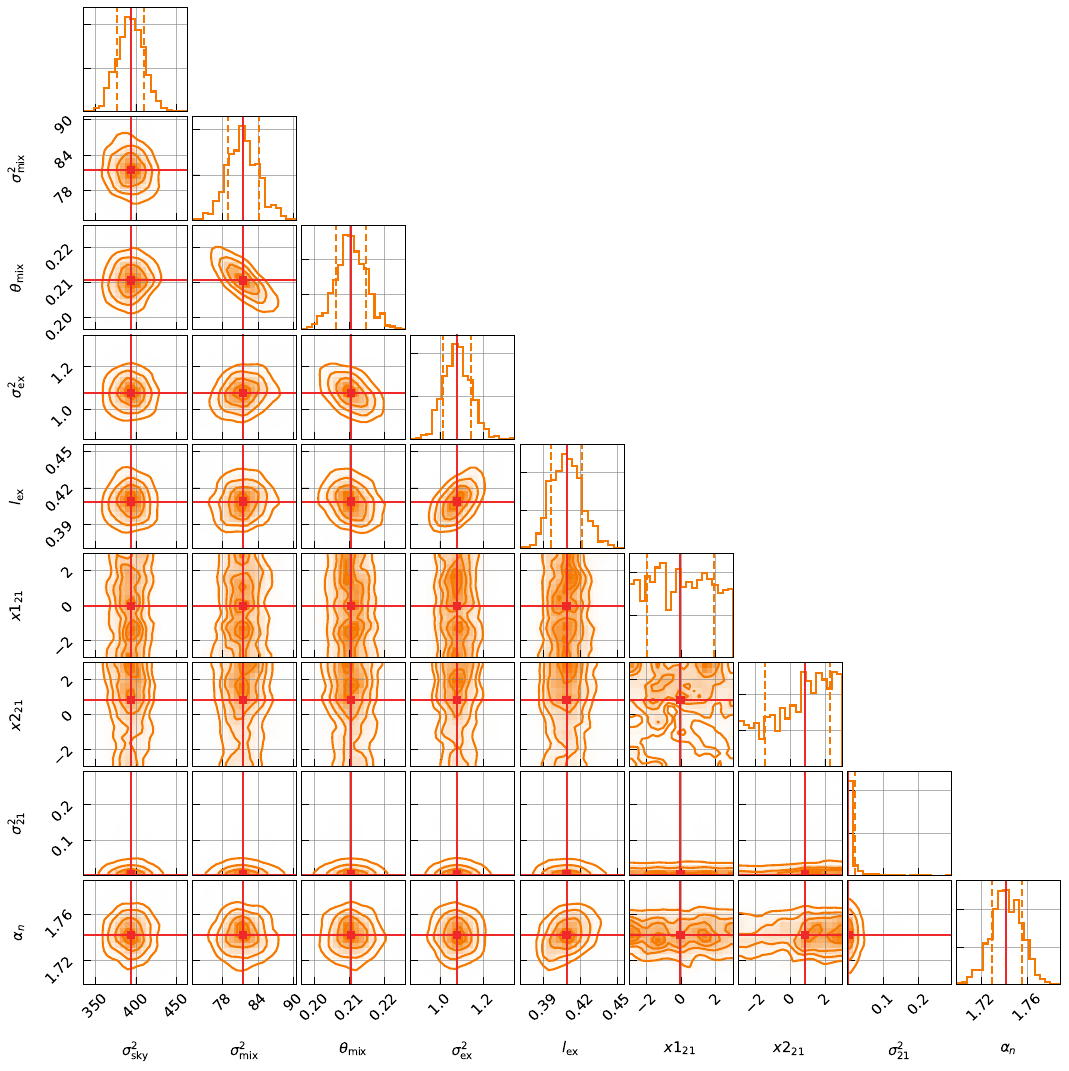}
    \caption{Posterior distribution of the Gaussian Process model parameters derived using a nested sampling algorithm, at $z \approx 8.3$}
    \label{fig:gpr_corner_eor3}
\end{figure*}

\end{document}